\documentclass[twocolumn,showpacs,preprintnumbers,amsmath,amssymb,prb]{revtex4}
\usepackage{graphicx}% Include figure files
\usepackage{dcolumn}% Align table columns on decimal point
\usepackage{bm}% bold math
\begin{document}

\title{Many-body corrections to ESR energy and spin-wave excitations in two-dimensional systems with Bychkov-Rashba spin splitting}% Force line breaks with \\

\author{S.~S.~Krishtopenko}
\affiliation{Institute for Physics of Microstructures RAS, Nizhny Novgorod, GSP-105, 603950, Russia}
\email[]{ds_a-teens@mail.ru}

\date{\today}% It is always \today, today,

\begin{abstract}
We report effects of electron-electron (\emph{e-e}) interaction on electron spin resonance (ESR) in perpendicular magnetic field in two-dimensional (2D) systems with Bychkov-Rashba spin splitting induced by spin-orbit interaction (SOI) and structural inversion asymmetry (SIA). Using the Hartree-Fock approximation, we demonstrate that the SIA results in non-zero many-body corrections to the ESR energy and the energy of spin wave excitations. We discover that the \emph{e-e} interaction in 2D systems with SIA not only can enhance the ESR energy but can also lead to the ESR energy reduction. The magnitude of this effect exhibits remarkable features in a wide range of parameters relevant to experiment: it is found to be rather sensitive to the sign of g-factor and the filling factor of Landau levels $\nu$. We derive analytical expressions for many-body corrections to ESR energy and the dispersion of spin wave excitations for the case of $\nu\leq 2$. We have found out that \emph{e-e} interaction does not affect the ESR energy in the case of filling of the lowest Landau level ($\nu\leq 1$) in 2D systems with positive g-factors even at arbitrarily large values of Bychkov-Rashba constant. The many-body renormalization of ESR energy in the case of fractional Quantum Hall effect is also discussed.
\end{abstract}

\pacs{73.21.Fg, 73.43.Lp, 73.61.Ey, 75.30.Ds, 75.70.Tj, 76.60.-k} % PACS, the Physics and Astronomy
\keywords{2D system, spin resonance, Larmor theorem, electron-electron interaction, Bychkov-Rashba spin splitting}
%Use showkeys class option if keyword                            %display desired
\maketitle

\section{\label{sec:Introduction}Introduction}
The investigation of the behavior of collective spin-wave (SW) excitations in two-dimensional (2D) systems can provide deep insights into the nature of its long-range magnetic order\cite{q1}. According to Larmor theorem\cite{q2}, in 2D systems with continuous rotational invariance in the spin space the long-wavelength collective SW excitation occurs exactly at the single-particle electron spin resonance (ESR) energy. In other words, electron-electron (\emph{e-e}) interaction does not contribute in the ESR energy in such systems.

The presence of SOI perturbs the spin invariance in the system, as well as contributes in the various \emph{e-e} interaction-induced effects in 2D electron gas (2DEG) in the integer\cite{q3,q4,q5,q6,q7,q8,q9,q10,q11} and fractional\cite{q12,q13,q14,q15,q16,q17} Quantum Hall effect (FQHE) regime. Controlling of the many-body effects via SOI is interesting from the fundamental physics point of view and is possibly useful for various device applications. This realization in semiconductor heterostructures through the Bychkov-Rashba (BR) spin splitting\cite{q18}, which arises from structural inversion asymmetry (SIA) in the presence of SOI, is of particular interest to spintronics research, because SOI-strength can be tuned by external gate voltages\cite{q19,q20,q21} or via persistent photoconductivity effect\cite{q22,q23,q24}.

The first theoretical evidence of the SOI-induced violation of the Larmor theorem in a 2DEG was reported by Califano et al.\cite{q12}, who studied effect of BR term, being linear in quasimomentum\cite{q18}, on many-body corrections to the ESR energy in the FQHE regime. Recently, we have provided a first theoretical evidence of the Larmor theorem violation in symmetric narrow-gap quantum wells (QWs)\cite{q10,q17}. We have shown that the ESR energy is significantly enhanced by \emph{e-e} interaction due to both SOI and the mixing of $\Gamma_6$ band with the bands $\Gamma_7$ and $\Gamma_8$. To describe the single-electron states in such narrow-gap QWs, the 8$\times$8 \textbf{k$\cdot$p} Hamiltonian was used.

Our latest paper\cite{q11} is devoted to many-body renormalization of the ESR energy and spin-wave excitations in symmetric and asymmetric QW based on narrow-gap materials. Since the inclusion of any asymmetric electric field in the 8-band \textbf{k$\cdot$p} Hamiltonian automatically leads to the effect of BR spin splitting in 2D system\cite{q25,q26}, we can, by comparing the calculations for the symmetric and asymmetric QW, evaluate the BR effect on many-body renormalization of the ESR energy in a narrow-gap QW. Note that by using 8-band \textbf{k$\cdot$p} Hamiltonian for single-electron states, we directly took into account the strong mixing between the conduction ($\Gamma_6$) and valence ($\Gamma_7$ and $\Gamma_8$) bands, which plays a principal role in many-body effects\cite{q4,q5,q6,q9,q10,q11,q17,q60} and results in the subband nonparabolicity in narrow-gap QWs. The nonparabolicity, in its turn, leads not only to the energy dependence of the effective mass\cite{q27,q28}, as measured in the cyclotron resonance experiments\cite{q29,q30,q31}, but to a nonlinear dependence of BR spin splitting on quasimomentum as well\cite{q25,q26}. The results reported in Ref.~\onlinecite{q11} demonstrate that BR spin splitting in asymmetric narrow-gap QWs leads to additional enhancement of many-particle ESR energy as compared with the many-particle values in the symmetric QWs.

In many semiconductor 2D systems one may neglect the subband nonparabolicity and describe the energy spectrum by using just a 'single-band' approximation. Within this model the electron mass and g-factor are assumed independent of energy and SOI in the presence of SIA is described by an additional BR term\cite{q18}, which is linear in quasimomentum. This work is devoted to studies of the \emph{e-e} interaction effects in ESR in 2DEG with BR spin splitting, placed in a perpendicular magnetic field, and to calculation of the SW dispersions in such system in the absence of nonparabolicity and disorder.

Surprisingly, we discover that the \emph{e-e} interaction in 2D systems with SIA can not only enhance the ESR energy, as predicted for narrow-gap QWs\cite{q11}, but can also can lead to the ESR energy \emph{reduction}. The magnitude of this effect is found to be rather sensitive to the sign of g-factor and the filling factor of Landau levels (LLs). Moreover, \emph{e-e} interaction does not affect the ESR energy in 2D systems with positive g-factors in the case of the lowest LL filling even at arbitrarily large values of BR constant. In particular, it allows one to obtain the single-particle g-factor values from ESR measurements in such 2D systems. It could also be important for new classes of 2D systems, based on the chemical elements with high atomic numbers, such as Te- or I-terminated surfaces of BiTeI, BiTeCl or BiTeBr\cite{q32,q33}, in which the BR constant reaches values as high as 4-5 eV$\cdot${\AA}.

The paper is organized as follows. The general theory based on the Hartree-Fock approximation (HFA) for calculation of SW excitations and 'many-particle' ESR energies in 2DEG systems with SIA is given in Section~\ref{sec:Theory}. To demonstrate the theoretical results obtained, we perform calculations with the parameters, relevant for 2D systems based on GaAs/AlGaAs because the sign of electron g-factor in such 2D systems can be varied\cite{q34,q36}. We understand that in addition to the SOI-induced term caused by SIA, the term related with Bulk Inversion Asymmetry (BIA)\cite{q37} in the Hamiltonian of such 2D systems does also exist. Moreover, the SIA and BIA terms could be comparable in magnitude\cite{q38,q39,q40}. Since our aim is to study the many-body renormalization of ESR energy that is related with SIA, we do not include the BIA term in the consideration. Calculations of the ESR energies and the SW dispersions for different values of electron g-factor values are performed in Section~\ref{sec:RnD}. The main results of this work are summarized in Section~\ref{sec:Summary}.

\section{\label{sec:Theory} Theoretical formalism}
We consider a 2DEG at zero temperature in the plane ($x$, $y$) placed in a perpendicular magnetic field in the presence of BR SOI-induced term\cite{q18} and in the absence of disorder. For simplicity, the electron motion along the $z$ direction is neglected, and, thus we consider a purely 2D system with a zero width along the $z$ axis.

\subsection{Single-electron picture}
In the simplest form of 'single-band' approximation a Hamiltonian for single-electron states is of the form:
\begin{equation}
\label{eq:1}
H_{(1e)}=\dfrac{\hbar^{2}\left(k_{x}^{2}+k_{y}^{2}\right)}{2m^{*}}+\dfrac{1}{2}g^{*}\mu_{B}B\sigma_{z}+\alpha\left(k_{y}\sigma_{x}-k_{x}\sigma_{y}\right)
\end{equation}
where $B$ is the magnetic field strength, $m^{*}$ is the effective electron mass, $g^{*}$ is effective g-factor, $\mu_{B}>0$ is the Bohr magneton, $\sigma_{x}$, $\sigma_{y}$, $\sigma_{z}$ are the Pauli matrices, $\alpha$ is a Bychkov-Rashba constant and
\begin{eqnarray}
\label{eq:2}
k_{x}=-i\frac{\partial}{\partial x}+\frac{e}{\hbar c}A_{x},\notag\\
k_{y}=-i\frac{\partial}{\partial y}+\frac{e}{\hbar c}A_{y},
\end{eqnarray}
where $e>0$ is the elementary charge, $A$ is the magnetic vector potential.

To calculate the LL energies and wave functions of single-electron states, it is convenient to introduce the LL ladder operators as follows\cite{q4}:
\begin{eqnarray*}
b^{+}=\frac{a_{B}}{\sqrt{2}}(k_{x}+ik_{y}),\\
b=\frac{a_{B}}{\sqrt{2}}(k_{x}-ik_{y}),\\
bb^{+}-b^{+}b=1,
\end{eqnarray*}
where $a_{B}=\sqrt{\hbar c/eB}$ is the magnetic length. As a result, $H_{(1e)}$ can be written in the following form:
\begin{equation}
\label{eq:3}
H_{(1e)}=\hbar\omega_{c}\left(b^{+}b+1/2\right)+\dfrac{1}{2}g^{*}\mu_{B}B\sigma_{z}+i\tilde{\alpha}\left(
\begin{matrix}
%1st row
0 & b \\
%2nd row
-b^{+} & 0
\end{matrix}
\right).
\end{equation}
Here $\hbar\omega_{c}=eB/m^{*}c$ and $\tilde{\alpha}=\sqrt{2}\alpha/a_{B}$.

By using the Landau gauge for the magnetic vector potential $A=(0,Bx,0)$, we can write the single-electron wave functions of $H_{(1e)}$ as

\begin{equation}
\label{eq:4}
\Psi_{n,k}(x,y)=A_{n}\left(
\begin{matrix}
%1st row
|n-1,k\rangle\\
%2nd row
0
\end{matrix}
\right)+B_{n}\left(
\begin{matrix}
%1st row
0 \\
%2nd row
|n,k\rangle
\end{matrix}
\right),
\end{equation}
where $n$ is the LL index, $k$ is the parameter for the degenerate states within the same LL in the Landau gauge and $|n,k\rangle$ are the normalized harmonic oscillator functions defined as
\begin{equation*}
|n,k\rangle=\begin{cases}
0,~n<0,\\[3pt]
\dfrac{\exp\left(iky\right)}{\sqrt{2^{n}n!\sqrt{\pi}a_{B}L}}H_{n}\left(\dfrac{\tilde{x}}{a_{B}}\right)\exp\left(\dfrac{-\tilde{x}^{2}}{2a_{B}^{2}}\right),~n\geq0,
\end{cases}\notag\\[8pt]
\end{equation*}
\begin{equation}
\label{eq:5}
\tilde{x}=x-ka_{B}^2.
\end{equation}
Here $L$ is the sample size along the $y$ axis, $H_{n}(x)$ are the Hermitian polynomials with number $n$.

The eigenvalues of $H_{(1e)}$ at $n>0$ have two branches:
\begin{gather}
E^{(a)}_{n}=\hbar\omega_{c}n+\sqrt{E^{2}_{0}+\tilde{\alpha}^{2}n},\notag\\
E^{(b)}_{n}=\hbar\omega_{c}n-\sqrt{E^{2}_{0}+\tilde{\alpha}^{2}n}.
\label{eq:6a}\end{gather}

\begin{figure*}
  \begin{minipage}{0.49\linewidth} %0.49\linewidth
\center{\Large{\textbf{(a)}} \includegraphics [width=1.06\columnwidth, keepaspectratio] {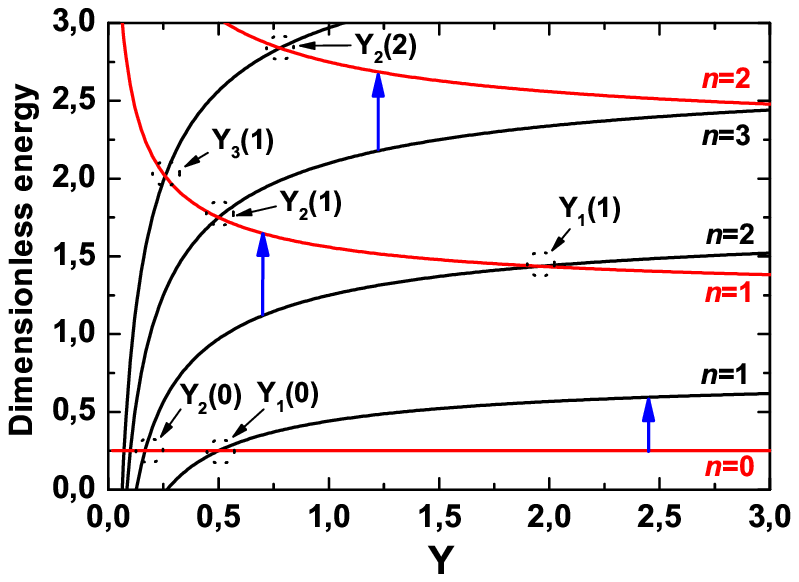}} % Here is how to import EPS art
   \end{minipage}
  \hfill
  \begin{minipage}{0.49\linewidth}
\center{\Large{\textbf{(b)}} \includegraphics [width=1.06\columnwidth, keepaspectratio] {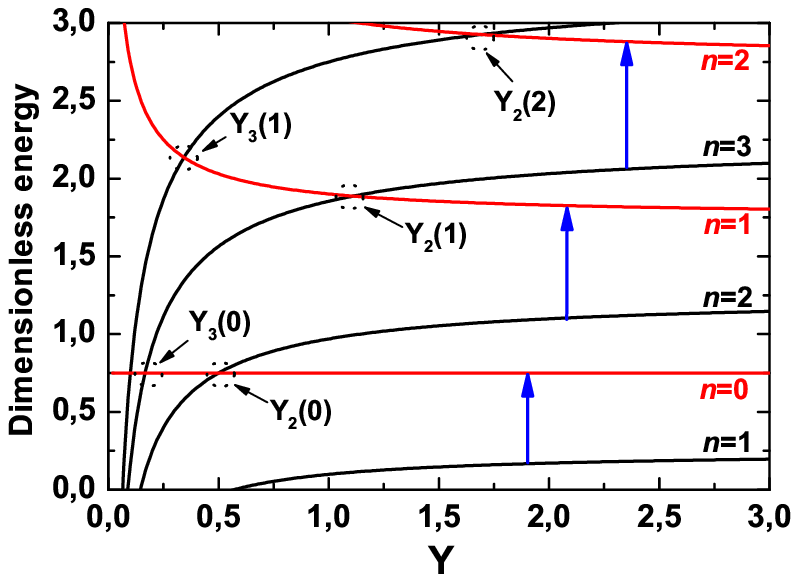}} % Here is how to import EPS art
  \end{minipage}
\caption{\label{Fig:1} (Color online) Dimensionless Landau levels for single-electron states as functions of dimensionless parameter $Y$ at (a) $A=0.5$ and (b) $A=1.5$.}
\end{figure*}

The single-electron wave functions of Hamiltonian~(\ref{eq:3}) are readily represented in the following form:
\begin{eqnarray}
\label{eq:7a}
\Psi_{n,k}^{(a)}=\sin\varphi_{n}\left(
\begin{matrix}
%1st row
0 \\
%2nd row
|n,k\rangle
\end{matrix}
\right)+i\cos\varphi_{n}\left(
\begin{matrix}
%1st row
|n-1,k\rangle\\
%2nd row
0
\end{matrix}
\right),
\end{eqnarray}
\begin{eqnarray}
\label{eq:7b}
\Psi_{n,k}^{(b)}=\sin\varphi_{n}\left(
\begin{matrix}
%1st row
|n-1,k\rangle\\
%2nd row
0
\end{matrix}
\right)+i\cos\varphi_{n}\left(
\begin{matrix}
%1st row
0 \\
%2nd row
|n,k\rangle
\end{matrix}
\right),
\end{eqnarray}
where
\begin{eqnarray}
\label{eq:8}
\sin\varphi_{n}=\dfrac{E_{0}+\sqrt{E^{2}_{0}+\tilde{\alpha}^{2}n}}{\sqrt{\left(E_{0}+\sqrt{E^{2}_{0}+\tilde{\alpha}^{2}n} \right)^2+\tilde{\alpha}^{2}n}},\notag\\
\cos\varphi_{n}=\dfrac{\tilde{\alpha}\sqrt{n}}{\sqrt{\left(E_{0}+\sqrt{E^{2}_{0}+\tilde{\alpha}^{2}n} \right)^2+\tilde{\alpha}^{2}n}}.
\end{eqnarray}

At $n=0$ the Hamiltonian~\eqref{eq:3} has only one eigenvalue:
\begin{equation}
\label{eq:9}
E_{0}=\dfrac{1}{2}\left(\hbar\omega_{c}-g^{*}\mu_{B}B\right),
\end{equation}
that corresponds to the wave function of the form
\begin{equation}
\label{eq:10}
\Psi_{0,k}=\left(
\begin{matrix}
%1st row
0 \\
%2nd row
|0,k\rangle
\end{matrix}
\right).
\end{equation}

Using the model proposed by Pfeffer and Zawadzki\cite{q36,q41}, it can be shown that relation
\begin{equation}
\label{eq:11}
\left|g^{*}\right|m^{*}<2m_{0},
\end{equation}
where $m_{0}$ is the free-electron mass, holds for the majority of 2D systems based on various semiconductor materials. It should be noted that condition~\eqref{eq:11} corresponds to $\hbar\omega_{c}>\left|g^{*}\right|\mu_{B}B$. Further we will restrict our consideration solely to such kind of 2D systems. Since, given condition~\eqref{eq:11} is met, eigenvalue~\eqref{eq:9} and wave function~\eqref{eq:10} can formally be derived from expressions~\eqref{eq:6a} and \eqref{eq:7a}, we will conditionally relate the solution with $n=0$ to the spectral branch labeled by index $a$.

For further analysis it is convenient to introduce the dimensionless LL energies as
\begin{eqnarray}
\label{eq:12}
E^{(a)}_{n}/\hbar\omega_{c}=n+\dfrac{1}{2}\sqrt{A^{2}+\dfrac{n}{Y}},\notag\\
E^{(b)}_{n}/\hbar\omega_{c}=n-\dfrac{1}{2}\sqrt{A^{2}+\dfrac{n}{Y}}.
\end{eqnarray}
where
\begin{eqnarray}
\label{eq:13}
A=1-\dfrac{g^{*}m^{*}}{2m_{0}},\notag\\
Y=4\left(\dfrac{\hbar\omega_{c}}{\tilde{\alpha}}\right)^{2}.
\end{eqnarray}
It is easily seen that relation~\eqref{eq:11} leads to condition $0<A<2$, with $A\geq 1$ corresponding to $g^{*}\leq 0$.

The dimensionless energy levels for single-electron states at $A=0.5$ ($g^{*}>0$) and $A=1.5$ ($g^{*}<0$) as functions of dimensionless parameter $Y$ are plotted in Fig.~\ref{Fig:1}. Red curves correspond to LLs $E_n^{(a)}/\hbar\omega_{c}$, black curves to the LL of the spectral branch described by $E_n^{(b)}/\hbar\omega_{c}$. The transitions between LLs ($n$,~$a$) and ($n+1$,~$b$), corresponding to ESR are indicated by blue arrows.

In addition to ESR, transitions corresponding to cyclotron resonance (($n$,~$a$)$\rightarrow$($n+1$,~$a$) and ($n$,~$b$)$\rightarrow$($n+1$,~$b$)) and combined resonance (($n$,~$a$)$\rightarrow$($n+2$,~$b$) and ($n$,~$b$)$\rightarrow$($n$,~$a$)) could be excited in 2D systems by external time-dependent magnetic or electric field. The excitation mechanism of these transitions was discussed in details by Rashba et al.\cite{q61,q62,q63} In this paper we focus on consideration the case of ESR only, therefore discussion of others electron transitions will be omitted.

An interesting feature of a 2D system with the BR spin splitting is the crossing of LLs from different branches as $Y$ changes with a varying $B$ or $\alpha$. One can easily show that the values of dimensionless parameter $Y_{k}(n)$, corresponding to the crossing of levels ($n$,~$a$) and ($n+k$,~$b$), satisfy the expression
\begin{equation}
\label{eq:14}
Y_{k}(n)=\dfrac{2n+k+\sqrt{4n(n+k)+A^2}}{k^2-A^2},
\end{equation}
It follows from Eq.~\eqref{eq:14} and condition $Y\geq 0$ that LLs ($n$,~$a$) and ($n+1$,~$b$) cross only if $A<1$, i.e., at $g^{*}>0$ . If this LL crossing occurs at the Fermi level, the single-electron values of ESR energy vanish. We note that condition $Y=Y_{1}(n)$ also gives rise to the divergency of spin Hall conductance\cite{q42}.

\begin{figure}
\includegraphics [width=1.08\columnwidth, keepaspectratio] {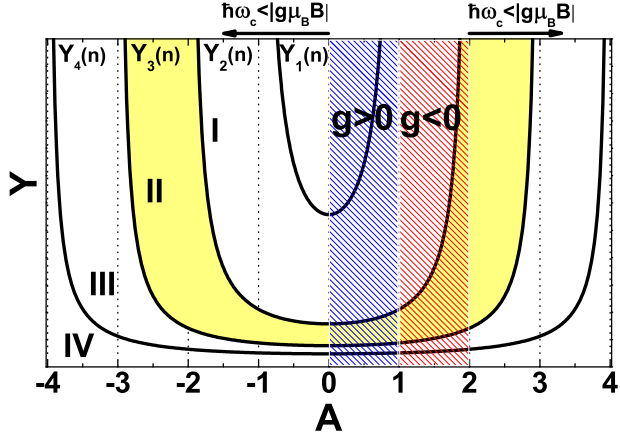}% Here is how to import EPS art
\caption{\label{Fig:2} (Color online) Diagram illustrating the regions of parameters corresponding to the varying number of ESR-related transitions in a 2D system with BR spin splitting at fixed values of LL number $n$.}
\end{figure}

Another important feature of a 2D system with BR spin splitting is that for certain relations between $m^{*}$, $g^{*}$, $\alpha$ and $B$ it simultaneously has several transitions associated with ESR. The diagram in Fig.~\ref{Fig:2} illustrates the regions of parameters corresponding to the varying number of the ESR-related transitions. At $Y_3(n)<Y<Y_2(n)$ (region II in Fig.~\ref{Fig:2}) for certain values of the LL filling factor there arises, besides the transition between levels ($n$,~$a$) and ($n+1$,~$b$), a transition between LLs ($n+1$,~$a$) and ($n+2$,~$b$), as well. Likewise, at $Y_4(n)<Y<Y_3(n)$ (region III in Fig.~\ref{Fig:2}) as many as three ESR transitions may arise simultaneously between following pairs of the LLs: ($n$,~$a$) and ($n+1$,~$b$); ($n+1$,~$a$) and ($n+2$,~$b$); ($n+2$,~$a$) and ($n+3$,~$b$). If $Y_{k+1}(n)<Y<Y_k(n)$, there is a possibility of $k$ transitions arising between pairs: ($n$,~$a$) and ($n+1$,~$b$); \ldots\ ($n+k-1$,~$a$) and ($n+k$,~$b$).

In weak magnetic fields that correspond to small values of $Y$ and large values of $n$ the number of the ESR transitions $k$ is also great. However, these transitions have close energies\cite{q43}:
\begin{gather}
\Delta_{ESR}^{(0)}=\lim_{n\rightarrow\infty}\left|E_n^{(a)}-E_{n+1}^{(b)}\right|\approx\notag\\ \approx\left|\left[(\hbar\omega_c-g^{*}\mu_BB)^2+\Delta_R^2\right]^{1/2}-\hbar\omega_c\right|,
\label{eq:15}\end{gather}
where the Fermi wave vector $k_F$ and BR spin splitting $\Delta_R$ in a zero magnetic field are defined as
\begin{eqnarray*}
\Delta_R=2\alpha k_F,\notag\\
k_F^2\approx\dfrac{2n}{a_B^2}.
\end{eqnarray*}

As mentioned above, we consider only the 2D systems in which the condition~\eqref{eq:11} is fulfilled, i.e., leads to $0<A<2$ (blue- and red-shaded areas). Besides, we will further examine the case of rather strong magnetic fields, $Y>Y_2(n)$, that in Fig.~\ref{Fig:2} corresponds to region I in which the single ESR-related transition is possible at fixed values of $n$.

\subsection{Hartree-Fock approximation}
The total Hamiltonian of the 2DEG with BR term in the second quantized representation can be written as:
\begin{gather}
\label{eq:16}
\hat{H}=\hat{H}_{0}+\hat{H}_{int},\notag\\
\hat{H}_{0}=\int d^2\vec{r}\hat{\Psi}^{+}(\vec{r})\hat{H}_{(1e)}\hat{\Psi}(\vec{r}),\notag\\
\hat{H}_{int}= \dfrac{1}{2}\int d^2\vec{r}_1\int d^2\vec{r}_2\hat{\Psi}^{+}(\vec{r}_1)\hat{\Psi}^{+}(\vec{r}_2)V(\left|\vec{r}_1-\vec{r}_2\right|)\times\notag\\
\times \hat{\Psi}(\vec{r}_2)\hat{\Psi}(\vec{r}_1),
\end{gather}
where $\vec{r}=(x,y)$ is a radius vector in 2DEG plane, the term $\hat{H}_{int}$ describes the \emph{e-e} interaction in 2DEG, $V(\left|\vec{r}_1-\vec{r}_2\right|)$ is the Coulomb potential, the upper sign '+' denotes the Hermitian conjugation. Here we have introduced the field operators $\hat{\Psi}(\vec{r})$ and $\hat{\Psi}^{+}(\vec{r})$, which are defined by fermion creation and annihilation operators $a_{n,k,i}$, $a^{+}_{n,k,i}$ and the single electron wave functions $\Psi_{n,k}^{(a)}$~(\ref{eq:7a}) and $\Psi_{n,k}^{(b)}$~(\ref{eq:7b}):
\begin{gather}
\label{eq:17}
\hat{\Psi}(\vec{r})=\sum_{n,k,i}\Psi_{n,k}^{(i)}(\vec{r})a_{n,k,i},\notag\\
\hat{\Psi}^{+}(\vec{r})=\sum_{n,k,i}\left(\Psi_{n,k}^{(i)}(\vec{r})\right)^{+}a^{+}_{n,k,i},
\end{gather}
where $i=a$ or $b$.

Using the Fourier transform for the Coulomb potential,
\begin{equation}
\label{eq:18}
V(|\vec{r}_1-\vec{r}_2|)=\int\frac{d^{2}\vec{q}}{(2\pi)^{2}}\tilde{D}(q)e^{i\vec{q}(\vec{r}_1-\vec{r}_2)},
\end{equation}
we can reduce the procedure of estimating the Coulomb potential matrix elements via the wave functions (\ref{eq:7a}) and (\ref{eq:7b}) to calculations of matrix elements $\langle n_{1},k_{1}|e^{i\vec{q}\vec{r}}|n_{2},k_{2}\rangle$\cite{q10,q17}.

The Fourier components of the Coulomb potential $\tilde{D}(q)$ for 2DEG, in general, have the form:
\begin{equation}
\label{eq:20}
\tilde{D}(q)=\dfrac{2\pi e^2}{\epsilon q}F(q),
\end{equation}
where $\epsilon$ is a permittivity of 2D system and $F(q)$ is geometrical form factor\cite{q44,q45} taking into account the nonzero thickness and electrostatic image force effects. As we have neglected the electron motion along the $z$ direction, we should set $F(q)=1$.

After some tedious calculations one arrives at the following expressions for $\hat{H}_{0}$ and $\hat{H}_{int}$:
\begin{gather}
\label{eq:21}
\hat{H}_{0}=\sum_{n,k,i} E_{n}^{(i)}a^{+}_{n,k,i}a_{n,k,i},\notag\\
\hat{H}_{int}= \dfrac{1}{2}\sum_{n_{1}...n_{4}}\sum_{i_{1}...i_{4}}\sum_{k_{1},k_{2}} \int \frac{d^2\vec{q}}{(2\pi)^2}\tilde{V}^{(i_{1},i_{2},i_{3},i_{4})}_{n_{1},n_{2},n_{3},n_{4}}(\vec{q})\times  \notag\\ \times e^{i q_{x}(k_{1}-k_{2}+q_{y})a_{B}^{2}}\times \notag\\ \times a^{+}_{n_1,k_1,i_1}a^{+}_{n_2,k_2,i_2}a_{n_3,k_2-q_y,i_3}a_{n_4,k_1+q_y,i_4},
\end{gather}
where the matrix elements $\tilde{V}^{(i_{1},i_{2},i_{3},i_{4})}_{n_{1},n_{2},n_{3},n_{4}}(\vec{q})$ are defined as
\begin{equation*}
\label{eq:22}
\tilde{V}^{(i_{1},i_{2},i_{3},i_{4})}_{n_{1},n_{2},n_{3},n_{4}}(\vec{q})=\tilde{D}(q)e^{-q^{2}a_{B}^{2}/2} \tilde{G}^{(i_{1},i_{4})}_{n_{1},n_{4}}(\vec{q})\tilde{G}^{(i_{2},i_{3})}_{n_{2},n_{3}}(-\vec{q}).
\end{equation*}
Here we have introduced $\tilde{G}^{(i_{1},i_{2})}_{n_{1},n_{2}}(\vec{q})$ as follows:
\begin{multline}
\label{eq:23}
\tilde{G}^{(i_1,i_2)}_{n_1,n_2}(\vec{q})= \tilde{L}^{(i_1,i_2)}_{n_1,n_2}\left(\frac{q^{2}a_{B}^{2}}{2}\right)\times\\
\times
\begin{cases}
\left[\dfrac{(iq_x+q_y)a_B}{\sqrt{2}}\right]^{n_1-n_2},~n_1\geq n_2,\\[3pt]
\left[\dfrac{(iq_x-q_y)a_B}{\sqrt{2}}\right]^{n_2-n_1},~n_1<n_2.
\end{cases}
\end{multline}
In Eq.~(\ref{eq:23}) $\tilde{L}^{(i_1,i_2)}_{n_1,n_2}\left(x\right)$, determined by wave functions (\ref{eq:7a}) and (\ref{eq:7b}), is of the following form:
\begin{gather}
\tilde{L}_{n,n'}^{(a,a)}\left(x\right)=\sin\varphi_{n}\sin\varphi_{n'}\sqrt{\dfrac{\tilde{n}_{1}!}{\tilde{n}_{2}!}}L_{\tilde{n}_{1}}^{\tilde{n}_{2}-\tilde{n}_{1}}\left(x\right)+\notag\\ +\cos\varphi_{n}\cos\varphi_{n'}\sqrt{\dfrac{(\tilde{n}_{1}-1)!}{(\tilde{n}_{2}-1)!}}L_{\tilde{n}_{1}-1}^{\tilde{n}_{2}-\tilde{n}_{1}}\left(x\right),\notag\\
\tilde{L}_{n,n'}^{(b,b)}\left(x\right)=\sin\varphi_{n}\sin\varphi_{n'}\sqrt{\dfrac{(\tilde{n}_{1}-1)!}{(\tilde{n}_{2}-1)!}}L_{\tilde{n}_{1}-1}^{\tilde{n}_{2}-\tilde{n}_{1}}\left(x\right)+\notag\\ +\cos\varphi_{n}\cos\varphi_{n'}\sqrt{\dfrac{\tilde{n}_{1}!}{\tilde{n}_{2}!}}L_{\tilde{n}_{1}}^{\tilde{n}_{2}-\tilde{n}_{1}}\left(x\right),\notag\\
\tilde{L}_{n,n'}^{(a,b)}\left(x\right)=i\sin\varphi_{n}\cos\varphi_{n'}\sqrt{\dfrac{\tilde{n}_{1}!}{\tilde{n}_{2}!}}L_{\tilde{n}_{1}}^{\tilde{n}_{2}-\tilde{n}_{1}}\left(x\right)-\notag\\ -i\cos\varphi_{n}\sin\varphi_{n'}\sqrt{\dfrac{(\tilde{n}_{1}-1)!}{(\tilde{n}_{2}-1)!}}L_{\tilde{n}_{1}-1}^{\tilde{n}_{2}-\tilde{n}_{1}}\left(x\right),
\label{eq:24a}\end{gather}
for $n>0$, $n'>0$, and
\begin{gather}
\tilde{L}_{0,n'}^{(a,a)}\left(x\right)=\dfrac{\sin\varphi_{n'}}{\sqrt{n'!}}, \notag\\ \tilde{L}_{0,n'}^{(a,b)}\left(x\right)=i\dfrac{\cos\varphi_{n'}}{\sqrt{n'!}}
\label{eq:24b}\end{gather}
for $n=0$, $n'>0$. Here $\tilde{n}_{1}=$min$(n,n')$, $\tilde{n}_{2}=$max$(n,n')$ and $L_{m}^{n}(x)$ are the associated Laguerre polynomials. Note that $\tilde{L}_{0,0}^{(a,a)}\left(x\right)=1$ and $\tilde{L}_{n_1,n_2}^{(i_1,i_2)}(x)=\left(\tilde{L}_{n_2,n_1}^{(i_2,i_1)}(x)\right)^{*}$, where the asterisk '*' denotes the complex conjugation.

Various collective excitations can be regarded as excitation of magnetic excitons formed by an electron that is excited onto an unfilled or partially filled LL ($n$, $i$), and an effective hole appearing simultaneously at level ($n'$, $i'$) abandoned by that electron\cite{q46}. To calculate the energies of such excitations it is convenient to use excitonic representation\cite{q10,q11,q47,q48,q49,q50}. Restricting our analysis to the zero temperature, let us introduce the exciton creation operator:
\begin{equation}
\label{eq:25}
A_{n,n',i,i'}^{+}(\vec{k})=\sum_{p}e^{ik_x(p+k_y/2)a^2_B}a^{+}_{n,p,i}a_{n',p+k_y,i'}.
\end{equation}
that satisfies the following commutation relation:
\begin{multline}
\label{eq:26}
\left[A_{n_1,n_2,i_1,i_2}^{+}(\vec{k}_1),A_{n_3,n_4,i_3,i_4}^{+}(\vec{k}_2)\right]=e^{-\frac{i}{2}a^2_{B}{[\vec{k}_1\times\vec{k}_2]}_z}\times\\
\times A_{n_1,n_4,i_1,i_4}^{+}(\vec{k}_1+\vec{k}_2)\delta_{n_2,n_3}\delta_{i_2,i_3}-\delta_{n_1,n_4}\delta_{i_1,i_4}\times\\\times
e^{\frac{i}{2}a^2_{B}{[\vec{k}_1\times\vec{k}_2]}_z}A_{n_3,n_2,i_3,i_2}^{+}(\vec{k}_1+\vec{k}_2).
\end{multline}
The excitation energy $E_{ex}$ with respect to the energy of the ground state $|0\rangle$ obeys the equation:
\begin{eqnarray}
\label{eq:27}
E_{ex}A_{n,n',i,i'}^{+}(\vec{k})|0\rangle=\left(E_{n}^{(i)}-E_{n'}^{(i')}\right)A_{n,n',i,i'}^{+}(\vec{k})|0\rangle+\notag\\ +\left[\hat{H}_{int},A_{n,n',i,i'}^{+}(\vec{k})\right]|0\rangle.~~~~~~~~~~~~~~~~~~~~~
\end{eqnarray}

To calculate the commutator in the right-hand side of Eq.~\eqref{eq:27}, involving six fermion operators, we have expressed $\hat{H}_{int}$ via $A_{n,n',i,i'}^{+}(\vec{k})|0\rangle$ as

\begin{multline}
\label{eq:28}
\hat{H}_{int}=\dfrac{1}{2}\sum_{\substack{n_{1}...n_{4}\\i_{1}...i_{4}}}\int \frac{d^2\vec{q}}{(2\pi)^2}\tilde{V}^{(i_{1},i_{2},i_{3},i_{4})}_{n_{1},n_{2},n_{3},n_{4}}(\vec{q})\times\\\times A_{n_1,n_4,i_1,i_4}^{+}(\vec{q})A_{n_2,n_3,i_2,i_3}^{+}(-\vec{q})-\\
-\dfrac{1}{2}\sum_{\substack{n_{1},n_{2},n_{3}\\i_{1},i_{2},i_{3}}}\int \frac{d^2\vec{q}}{(2\pi)^2}\tilde{V}^{(i_{1},i_{2},i_{2},i_{3})}_{n_{1},n_{2},n_{2},n_{3}}(\vec{q})A_{n_1,n_3,i_1,i_3}^{+}(0),
\end{multline}

Now using the commutation relations~\eqref{eq:26} and following the standard HFA rule
\begin{equation}
\label{eq:30}
\langle0|a_{n_1,k_1,i_1}a^{+}_{n_2,k_2,i_2}|0\rangle=\nu_{n_1}^{(i_1)}\delta_{n_1,n_2}\delta_{p_1,p_2}\delta_{i_1,i_2},
\end{equation}
where $\nu_{n}^{(i)}$ is the filling factor for LL ($n$, $i$), we get the following expression for the commutator in the right-hand part of Eq.~\eqref{eq:27}:

\begin{gather}
\left[\hat{H}_{int},A_{n,n',i,i'}^{+}(\vec{k})\right]|0\rangle=
-\sum_{n_{2},i_{2}}\nu_{n_2}^{(i_2)}\Bigr(\tilde{E}^{(i,i_2,i,i_{2})}_{n,n_2,n,n_2}(0)-\notag\\
-\tilde{E}^{(i',i_2,i',i_{2})}_{n',n_2,n',n_2}(0)\Bigl)A_{n,n',i,i'}^{+}(\vec{k})|0\rangle-\notag\\
-(\nu_{n}^{(i)}-\nu_{n'}^{(i')})\sum_{\substack{n_1,n_4\\i_1,i_4}}\frac{\tilde{V}^{(i_1,i',i,i_4)}_{n_1,n',n,n_4}(\vec{k})}{2\pi a_B^2}A_{n_1,n_4,i_1,i_4}^{+}(\vec{k})|0\rangle+\notag\\
+(\nu_{n}^{(i)}-\nu_{n'}^{(i')})\sum_{\substack{n_1,n_2\\i_1,i_2}}\tilde{E}^{(i',i_1,i,i_2)}_{n',n_1,n,n_2}(\vec{k})A_{n_1,n_2,i_1,i_2}^{+}(\vec{k})|0\rangle
\label{eq:31}\end{gather}
where
\begin{equation}
\label{eq:32}
\tilde{E}^{(i_1,i_2,i_3,i_4)}_{n_1,n_2,n_3,n_4}(\vec{k})=\int \frac{d^2\vec{q}}{(2\pi)^2}\tilde{V}^{(i_1,i_2,i_3,i_4)}_{n_1,n_2,n_3,n_4}(\vec{q})e^{i a^2_{B}{[\vec{q}\times\vec{k}]}_z}.
\end{equation}
As was done in our previous paper~\cite{q11}, of all the terms in the expression for $\left[\hat{H}_{int},A_{n,n',i,i'}^{+}(\vec{k})\right]$ we retain just those containing one creation- and one annihilation fermion operator, multiplied by the operator for the number of particles. This is fully equivalent to the 'mean-field' approach developed by Kallin and Halperin\cite{q46}. Eq.~\eqref{eq:31} formally coincides with that obtained in Ref.~\onlinecite{q11}. The difference is hidden in the calculation of the matrix elements of \emph{e-e} interaction. As clearly seen from Eq.~\eqref{eq:32}, the second and third terms provide the mixing of all possible excitation states in a 2D system.

Let us look into the details of the excitation induced by an electron transition between LLs ($n$,~$a$) and ($n-1$,~$b$), whose energy in the long-wave limit corresponds to that of ESR. In the absence of BR spin splitting ($\alpha=0$) this excitation corresponds to spin wave excitation (or spin exciton). Strictly speaking, spin is not a good quantum number for classification of the excitations at $\alpha\neq 0$. Nevertheless, as in our previous work\cite{q11}, we will adhere to the term 'spin wave' for the excitation between LL ($n$,~$a$) and ($n-1$,~$b$) in a 2D system with BR spin splitting, by analogy with the case $\alpha=0$. Note that electron transitions between the LLs ($n$,~$i$) and ($n+1$,~$i$) corresponds to the magnetoplasmon excitations, which energies at $k=0$ are measured in cyclotron resonance experiments.

As it is mentioned above, we restrict ourself by consideration the case of rather strong magnetic fields, which correspond to $Y>Y_2(n)$. We note that only single ESR-related transition is possible in this case. It is clear from Fig.~\ref{Fig:1} that in the vicinity of $Y_2(n)$ ESR energy is comparable with the energy of cyclotron resonance transition of an electron from LL ($n+1$,~$b$) to the level ($n+2$,~$b$). Therefore, the excitations described by operators $A_{n+2,n+1,b,b}^{+}(\vec{k})$ and $A_{n,n+1,a,b}^{+}(\vec{k})$ are mixed. However, the zero-momentum excitations corresponding to the cyclotron resonance and ESR do not interact due to $\tilde{V}^{(b,b,a,b)}_{n+2,n+1,n,n+1}(0)=0$ and $\tilde{E}^{(b,b,a,b)}_{n+1,n+2,n,n+1}(0)=0$. Further, we neglect the mixing between excitations $A_{n+2,n+1,b,b}^{+}(\vec{k})|0\rangle$ and $A_{n,n+1,a,b}^{+}(\vec{k})|0\rangle$, which is valid for $k=0$ at any values of $Y$ or for $Y\gg Y_2(n)$ at non-zero momentum values.

By taking into consideration the above arguments and by using the polar coordinate system to calculate the integral over the polar angle in the matrix elements~\eqref{eq:32}, one arrives at the expressions for SW excitation energy:
\begin{widetext}
\begin{gather}
E_{SW}(k)=\left|E_{n}^{(a)}-E_{n+1}^{(b)}+\dfrac{e^2}{\epsilon a_B}\Delta^{(e-e)}_{SW}(k)\right|,\notag\\
\Delta^{(e-e)}_{SW}(k)=\left(\nu_{n+1}^{(b)}-\nu_{n}^{(a)}\right)\dfrac{ka_B}{2}e^{-\dfrac{k^{2}a_{B}^{2}}{2}}F(k)\left|\tilde{L}_{n,n+1}^{(a,b)}\left(\dfrac{k^2a_B^2}{2}\right)\right|^2-\notag\\ -\left(\nu_{n+1}^{(b)}-\nu_{n}^{(a)}\right)\int\limits_{0}^{+\infty}dxe^{-\dfrac{x^{2}}{2}}F\left(\dfrac{x}{a_B}\right)J_0\left(ka_Bx\right)\tilde{L}_{n+1,n+1}^{(b,b)}\left(\dfrac{x^2}{2}\right)\tilde{L}_{n,n}^{(a,a)}\left(\dfrac{x^2}{2}\right)-\notag\\ -\nu_{n}^{(a)}\int\limits_{0}^{+\infty}dxe^{-\dfrac{x^{2}}{2}}F\left(\dfrac{x}{a_B}\right)\left\{\left|\tilde{L}_{n,n}^{(a,a)}\left(\dfrac{x^2}{2}\right)\right|^2-\dfrac{x^2}{2}\left|\tilde{L}_{n+1,n}^{(b,a)}\left(\dfrac{x^2}{2}\right)\right|^2\right\}-\notag\\ -\nu_{n+1}^{(b)}\int\limits_{0}^{+\infty}dxe^{-\dfrac{x^{2}}{2}}F\left(\dfrac{x}{a_B}\right)\left\{\dfrac{x^2}{2}\left|\tilde{L}_{n,n+1}^{(a,b)}\left(\dfrac{x^2}{2}\right)\right|^2-\left|\tilde{L}_{n+1,n+1}^{(b,b)}\left(\dfrac{x^2}{2}\right)\right|^2\right\}+\hat{\Sigma}_n^{(a)}-\hat{\Sigma}_{n+1}^{(b)}
\label{eq:32a}\end{gather}
for $Y\leq Y_1(n)$ if $g^{*}>0$ or for $g^{*}\leq 0$, and

\begin{equation*}
E_{SW}(k)=\left|E_{n+1}^{(b)}-E_{n}^{(a)}+\dfrac{e^2}{\epsilon a_B}\Delta^{(e-e)}_{SW}(k)\right|,
\end{equation*}
\begin{gather}
\Delta^{(e-e)}_{SW}(k)=\left(\nu_{n}^{(a)}-\nu_{n+1}^{(b)}\right)\dfrac{ka_B}{2}e^{-\dfrac{k^{2}a_{B}^{2}}{2}}F(k)\left|\tilde{L}_{n,n+1}^{(a,b)}\left(\dfrac{k^2a_B^2}{2}\right)\right|^2-\notag\\ -\left(\nu_{n}^{(a)}-\nu_{n+1}^{(b)}\right)\int\limits_{0}^{+\infty}dxe^{-\dfrac{x^{2}}{2}}F\left(\dfrac{x}{a_B}\right)J_0\left(ka_Bx\right)\tilde{L}_{n+1,n+1}^{(b,b)}\left(\dfrac{x^2}{2}\right)\tilde{L}_{n,n}^{(a,a)}\left(\dfrac{x^2}{2}\right)-\notag\\ -\nu_{n}^{(a)}\int\limits_{0}^{+\infty}dxe^{-\dfrac{x^{2}}{2}}F\left(\dfrac{x}{a_B}\right)\left\{\dfrac{x^2}{2}\left|\tilde{L}_{n+1,n}^{(b,a)}\left(\dfrac{x^2}{2}\right)\right|^2-\left|\tilde{L}_{n,n}^{(a,a)}\left(\dfrac{x^2}{2}\right)\right|^2\right\}-\notag\\ -\nu_{n+1}^{(b)}\int\limits_{0}^{+\infty}dxe^{-\dfrac{x^{2}}{2}}F\left(\dfrac{x}{a_B}\right)\left\{\left|\tilde{L}_{n+1,n+1}^{(b,b)}\left(\dfrac{x^2}{2}\right)\right|^2-\dfrac{x^2}{2}\left|\tilde{L}_{n,n+1}^{(a,b)}\left(\dfrac{x^2}{2}\right)\right|^2\right\}+\hat{\Sigma}_{n+1}^{(b)}-\hat{\Sigma}_n^{(a)}
\label{eq:32b}\end{gather}
for $Y>Y_1(n)$ if $g^{*}>0$.

In Eqs.~\eqref{eq:32a} and \eqref{eq:32b}, $J_0(x)$ is the zero-order Bessel function, $\hat{\Sigma}_{n+1}^{(b)}$ and $\hat{\Sigma}_n^{(a)}$ are the exchange contributions of \emph{completely occupied} LLs to the energy of LL ($n+1$,~$b$) and ($n$,~$a$). We note that $\hat{\Sigma}_{n+1}^{(b)}=0$ and $\hat{\Sigma}_n^{(a)}=0$ at $n=0$. At $n>0$ they are defined as
\begin{gather}
\hat{\Sigma}_n^{(a)}=-\sum_{n_{2}=0}^{n-1}\int\limits_{0}^{+\infty}dxe^{-\dfrac{x^{2}}{2}}\left(\dfrac{x^{2}}{2}\right)^{\left|n-n_2\right|}F\left(\dfrac{x}{a_B}\right)\left\{\left|\tilde{L}_{n,n_2}^{(a,a)}\left(\dfrac{x^2}{2}\right)\right|^2+\dfrac{x^2}{2}\left|\tilde{L}_{n,n_2+1}^{(a,b)}\left(\dfrac{x^2}{2}\right)\right|^2\right\},\notag\\
\hat{\Sigma}_{n+1}^{(b)}=-\sum_{n_{2}=0}^{n-1}\int\limits_{0}^{+\infty}dxe^{-\dfrac{x^{2}}{2}}\left(\dfrac{x^{2}}{2}\right)^{\left|n-n_2\right|}F\left(\dfrac{x}{a_B}\right)\left\{\left|\tilde{L}_{n+1,n_2+1}^{(b,b)}\left(\dfrac{x^2}{2}\right)\right|^2+\dfrac{x^2}{2}\left|\tilde{L}_{n_2,n+1}^{(a,b)}\left(\dfrac{x^2}{2}\right)\right|^2\right\}.
\label{eq:33}\end{gather}
\end{widetext}

Eqs~\eqref{eq:32a}, \eqref{eq:32b} and \eqref{eq:33} have a general form and are valid also in 2D systems with a nonzero thickness along the $z$ direction. It is seen that $\Delta^{(e-e)}_{SW}(k)$ is a finite quantity at $n\rightarrow\infty$. Since this case corresponds to the limit of weak magnetic fields at fixed values of the 2D electrons concentration, $e^2/(\epsilon a_B)\Delta^{(e-e)}_{SW}(k)\rightarrow 0$ at $B\rightarrow 0$, which implies that the ESR energy calculated within HFA tends to zero with a decreasing magnetic field.

\subsection{HFA in the case of $n = 0$}
Now consider a 2D system with $F(q)=1$, in which excitation occurs between levels ($0$,~$a$) and ($1$,~$b$), i.e., when the LL filling factor is $\nu\leq 2$. In this case the integrals in Eqs.~\eqref{eq:32a} and \eqref{eq:32b} can be calculated analytically. As a result, the expression for $\Delta^{(e-e)}_{SW}(k)$ can be written as
\begin{gather}
\Delta^{(e-e)}_{SW}(k)=\left|\nu_{0}^{(a)}-\nu_{1}^{(b)}\right|\Delta^{(e-e)}(k)+\notag\\ +\left|\nu_{0}^{(a)}-\nu_{1}^{(b)}\right|\Delta^{(e-e)}_{ab}(k)\pm\nu_{1}^{(b)}\Delta^{(e-e)}_{b}.
\label{eq:34b}\end{gather}
Here '$+$' corresponds to the case $E^{(b)}_1>E^{(a)}_0$, i.e., for $g^{*}>0$ and $Y>Y_1(0)$, while '$-$' matches the occasion $E^{(a)}_0>E^{(b)}_1$, which takes place if $g^{*}>0$ and $Y\leq Y_1(0)$ or if $g^{*}\leq 0$. The momentum-dependent contributions $\Delta^{(e-e)}(k)$ and $\Delta^{(e-e)}_{ab}(k)$ are defined as
\begin{equation}
\label{eq:35a}
\Delta^{(e-e)}(k)=\dfrac{ka_B}{2}e^{-k^{2}a_{B}^{2}/2}\cos^2\varphi_1,
\end{equation}
\begin{gather}
\Delta^{(e-e)}_{ab}(k)=-\sqrt{\dfrac{\pi}{2}}\dfrac{\cos^2\varphi_1}{2}\dfrac{k^2a_B^2}{2}e^{-k^{2}a_{B}^{2}/4}\times\notag\\ \times\left[I_0\left(\dfrac{k^{2}a_{B}^{2}}{4}\right)-I_1\left(\dfrac{k^{2}a_{B}^{2}}{4}\right)\right]+\notag\\ +\sqrt{\dfrac{\pi}{2}}\left\{1-\dfrac{\cos^2\varphi_1}{2}\right\}\left[1-I_0\left(\dfrac{k^{2}a_{B}^{2}}{4}\right)e^{-k^{2}a_{B}^{2}/4}\right],
\label{eq:35}\end{gather}
where $I_0(x)$ and $I_1(x)$ are the modified zero- and first-order Bessel functions.
In Eq.~\eqref{eq:34b} $\Delta^{(e-e)}_{b}$ does not depend on momentum and is written as follows:
\begin{equation}
\label{eq:36}
\Delta^{(e-e)}_{b}=\sqrt{\dfrac{\pi}{2}}\left(1-\dfrac{3}{4}\cos^2\varphi_1\right)\cos^2\varphi_1.
\end{equation}

At $\alpha\rightarrow 0$ ($\cos^2\varphi_1\rightarrow 0$) Eq.~\eqref{eq:34b} takes the form:
\begin{equation*}
\Delta^{(e-e)}_{SW}(k)=\left|\nu_{0}^{(a)}-\nu_{1}^{(b)}\right|\sqrt{\dfrac{\pi}{2}}\left[1-J_0\left(\frac{k^{2}a_{B}^{2}}{4}\right)e^{-k^2a_B^2/4}\right],
\end{equation*}
which has previously been obtained within HFA in the absence of BR term\cite{q46,q51,q52}.

The many-body correction to ESR energy is defined by nonvanishing terms at $k=0$ in Eq.~\eqref{eq:34b}:
\begin{equation}
\label{eq:37}
\Delta^{(e-e)}_{SW}(0)=\pm\nu_{1}^{(b)}\Delta^{(e-e)}_{b}.
\end{equation}
Eq.~\eqref{eq:37} being valid for any values of BR constant $\alpha$ has some remarkable features.
One can see that in 2D systems with negative g-factor values (the case of '$-$') at $\nu\leq2$ the correction is negative, i.e., the \emph{e-e} interaction causes \emph{reduction} of ESR energy.

In 2D systems with positive values of the electron g-factor in strong magnetic fields such that $Y>Y_1(0)$ (the case of '$+$') the filling factor increasing in the interval $1<\nu\leq2$ the correction is positive, i.e., the \emph{e-e} interaction leads to \emph{enhancement} of the ESR energy.

At $\nu\leq 1$ the \emph{e-e} interaction does not affect the ESR energy. Indeed, in these circumstances $E_1^{(b)}>E_0^{(a)}$, hence, at thermodynamical equilibrium $\nu_{1}^{(a)}=\nu$ and $\nu_{1}^{(b)}=0$. The latter condition provides that many-body correction to the ESR energy, $\Delta^{(e-e)}_{SW}(0)$, turns to zero at $\nu\leq1$ even at arbitrarily large values of BR constant. This results from a specific structure of electron wave function \eqref{eq:10} in the lowest LL, which is the same as it is in the absence of BR SOI-induced term. Further, we show that this remarkable behavior is not caused by HFA and also holds in non-perturbative approach, used for the case of FQHE.

\subsection{ESR at $\nu<1$ beyond HFA}
It is worth mentioning that HFA is a fairly good approximation for magnetoplasmon and spin-wave excitation at integer LL filling factors\cite{q46,q47,q48,q60}. At special (fractional) filling factors the ground state of 2D system is a highly correlated electron liquid (see, for example, Ref.~\onlinecite{q64}). The HFA, strictly speaking, does not take into account the correlated nature of the ground state at fractional LL filling factors. The electron correlations in the ground state can be taken into account in the generalized single-mode approximation (GSMA)\cite{q2,q17,q49,q65,q66}.

The energy of SW excitation within GSMA at $\nu<1$ has the form
\begin{equation}
\label{eq:38}
E_{SW}(\vec{k})=E_{n}^{(i)}-E_{n'}^{(i')}+\dfrac{e^2}{\epsilon a_B}\Delta^{(e-e)}_{SW}(\vec{k})
\end{equation}
where the contribution of \emph{e-e} interaction is determined as follows:
\begin{equation}
\label{eq:39}
\dfrac{e^2}{\epsilon a_B}\Delta^{(e-e)}_{SW}(\vec{k})=\frac{\langle0|A_{n',n,i',i}^{+}(-\vec{k})\left[H_{int},A_{n,n',i,i'}^{+} (\vec{k})\right]|0\rangle}{\langle0|A_{n',n,i',i}^{+}(-\vec{k})A_{n,n',i,i'}^{+}(\vec{k})|0\rangle}
\end{equation}
In Eq.~\eqref{eq:38} we implicitly consider ($n'$, $i'$) as the lowest LL. It can be shown that averaging of $\langle0|A_{n',n',i',i'}^{+}(-\vec{q}~)A_{n',n',i',i'}^{+}(\vec{q}~)|0\rangle$ over the ground state is proportional to the density-density correlation function calculated for the lowest LL~\cite{q66} that is, in its turn, also related to the pair distribution function $g_{n'}^{(i')}(r)$ for the electrons in LL ($n'$, $i'$) calculated for the ground state~\cite{q65,q66}:
\begin{gather}
\langle 0|A_{n',n',i',i'}^{+}(-\vec{q})A_{n',n',i',i'}^{+}(\vec{q})|0\rangle= \nu_{n'}^{(i')}N_\phi\left[s_{n'}^{(i')}(\vec{q}~)-1\right],\notag\\
s_{n'}^{(i')}(\vec{q}~)-1=\tilde{h}_{n'}^{(i')}(q)+2\pi\nu_{n'}^{(i')}\delta^2(\vec{q}~),
\label{eq:40}\end{gather}
where $s_{n'}^{(i')}(\vec{q}~)$ is a static structure factor for the electrons in the lowest LL, $N_\phi=L^2/2\pi a_B^2$ (where $L\times L$ is a square of 2D system) and $\tilde{h}_{n'}^{(i')}(q)$ is defined by the Fourier component of $g_{n'}^{(i')}(r)$:
\begin{equation}
\label{eq:41}
\tilde{h}_{n'}^{(i')}(q)=\nu_{n'}^{(i')}\int d^{2}\vec{r}\left[g_{n'}^{(i')}(r)-1\right]e^{-i\vec{q}\vec{r}}.
\end{equation}
In \eqref{eq:41} $g_{n'}^{(i')}(r)$ satisfies the normalization condition $\nu_{n'}^{(i')}\int d^{2}\vec{r}\left[g_{n'}^{(i')}(r)-1\right]=-1$. Note also that the exact equation $\langle 0|A_{n',n,i',i}^{+}(-\vec{k})A_{n,n',i,i'}^{+}(\vec{k})|0\rangle=\nu_{n'}^{(i')}N_\phi$ holds.

By using the commutation relations \eqref{eq:26} for the exciton creation operators and Eqs.~\eqref{eq:40}, it is straightforward to rewrite Eq.~\eqref{eq:39} in the form:
\begin{gather}
\dfrac{e^2}{\epsilon a_B}\Delta^{(e-e)}_{SW}(\vec{k})=\int \frac{d^2\vec{q}~}{(2\pi)^2}\tilde{V}^{(i,i',i,i')}_{n,n',n,n'}(\vec{q})\tilde{h}_{n'}^{(i')}(|\vec{k}-\vec{q}|)-\notag\\
-\int \frac{d^2\vec{q}~}{(2\pi)^2}\tilde{V}^{(i',i',i',i')}_{n',n',n',n'}(\vec{q}) \tilde{h}_{n'}^{(i')}(q)+\nu_{n'}^{(i')} \frac{\tilde{V}^{(i,i',i,i')}_{n,n',n,n'}(\vec{k})}{2\pi a_B^2}+\notag\\
+\int \frac{d^2\vec{q}~}{(2\pi)^2}\tilde{V}^{(i',i,i,i')}_{n',n,n,n'}(\vec{q})e^{ia^2_{B} {[\vec{q}\times\vec{k}]}_z}\tilde{h}_{n'}^{(i')}(q)
\label{eq:42}\end{gather}

First we consider the many-body correction to ESR energy in the case ($n'$, $i'$)=($0$, $a$) and ($n$, $i$)=($1$, $b$), i.e., at $g^{*}>0$ and $Y>Y_1(0)$. Under this condition Eq.~\eqref{eq:42} is rewritten as
\begin{gather}
\Delta^{(e-e)}_{SW}(0)=\int\limits_{0}^{+\infty}dxe^{-x^{2}/2}\tilde{h}_{0}^{(a)}\left(\dfrac{x}{a_B}\right)F\left(\dfrac{x}{a_B}\right)\times\notag\\ \times\Bigg\{\dfrac{x^2}{2}\left|\tilde{L}_{1,0}^{(b,a)}\left(\dfrac{x^2}{2}\right)\right|^2-\left|\tilde{L}_{0,0}^{(a,a)}\left(\dfrac{x^2}{2}\right)\right|^2+\tilde{L}_{1,1}^{(b,b)}\left(\dfrac{x^2}{2}\right)\Bigg\}.
\label{eq:43}\end{gather}
We note that since electron wave function for LL ($0$, $a$) is unaffected by SOI, $\tilde{h}_{0}^{(a)}(q)$ can be calculated from the parametrization of $g_{0}^{(a)}(r)$ proposed by Girvin\cite{q65,q66} for the Laughlin states (for LL filling factors $\nu_{0}^{(a)}=1/M$, where $M$ is an odd integer) in the absence of SOI.

Taking into account Eqs.~\eqref{eq:24a} and \eqref{eq:24b}, it is easy to verify that $\Delta^{(e-e)}_{SW}(0)$ in Eq.~\eqref{eq:43} is equal to zero. Thus, we have shown that even beyond the HFA \emph{e-e} interaction does not affect the ESR energy in 2D systems with positive g-factors at $Y>Y_1(0)$.

   \begin{figure*}
  \begin{minipage}{0.49\linewidth} %0.49\linewidth
\center{\Large{\textbf{(a)}} \includegraphics [width=1.07\columnwidth, keepaspectratio] {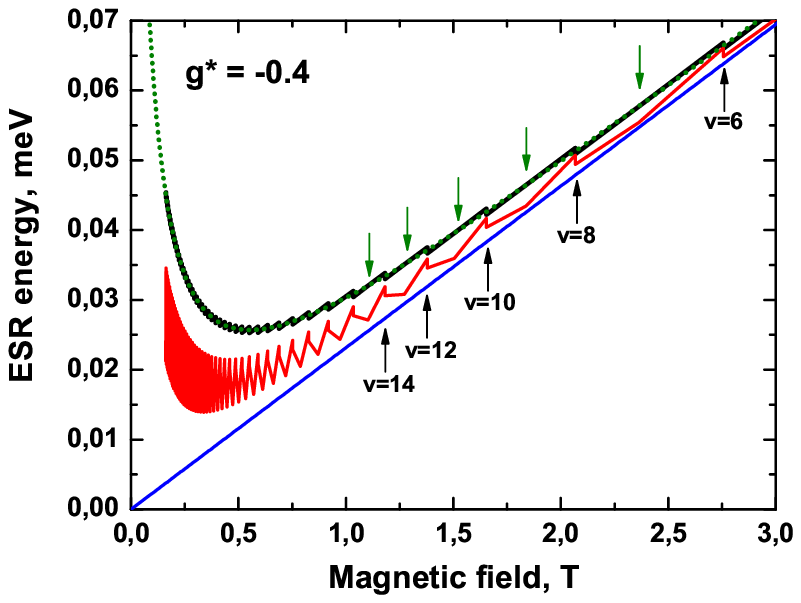}} % Here is how to import EPS art
   \end{minipage}
  \hfill
  \begin{minipage}{0.49\linewidth}
\center{\Large{\textbf{(b)}} \includegraphics [width=1.08\columnwidth, keepaspectratio] {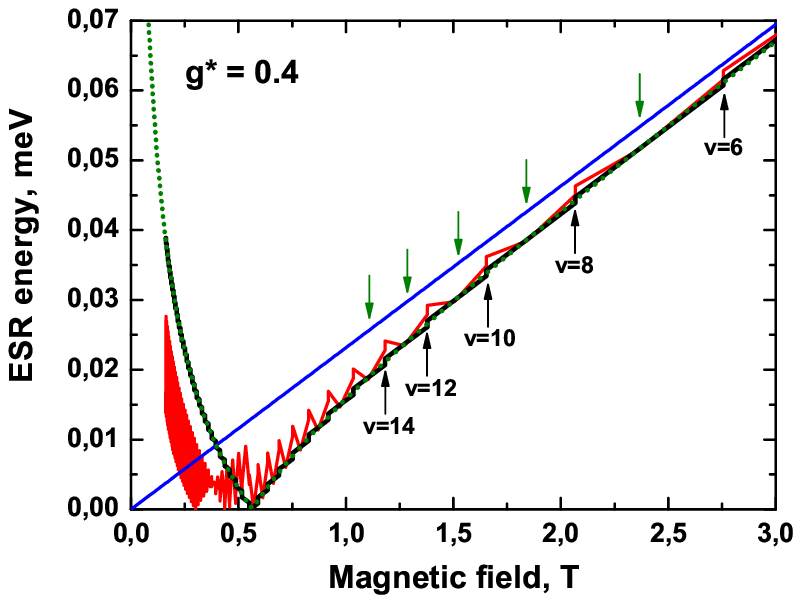}} % Here is how to import EPS art
  \end{minipage}
  \vfill
  \begin{minipage}{0.49\linewidth}
\center{\Large{\textbf{(c)}} \includegraphics [width=1.06\columnwidth, keepaspectratio] {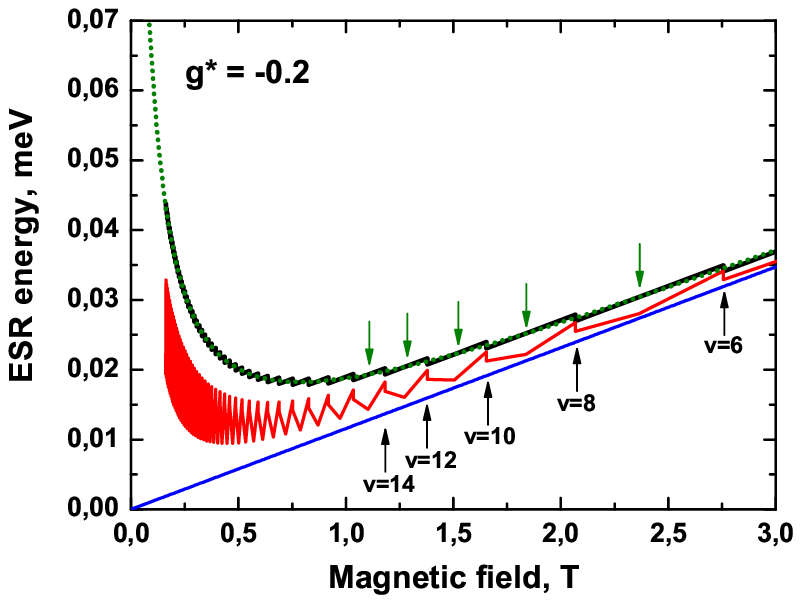}} % Here is how to import EPS art
  \end{minipage}
   \hfill
  \begin{minipage}{0.49\linewidth}
\center{\Large{\textbf{(d)}} \includegraphics [width=1.06\columnwidth, keepaspectratio] {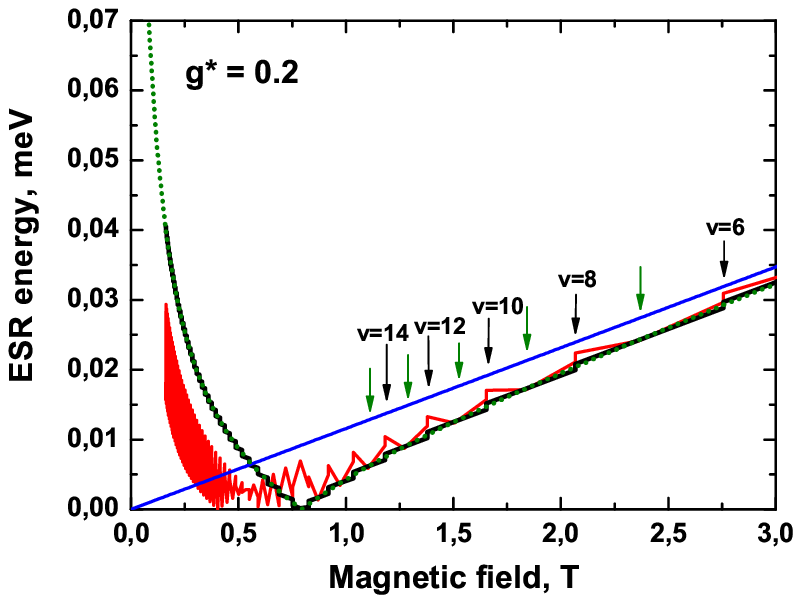}} % Here is how to import EPS art
  \end{minipage}
\vfill
  \begin{minipage}{0.49\linewidth}
\center{\Large{\textbf{(e)}} \includegraphics [width=1.07\columnwidth, keepaspectratio] {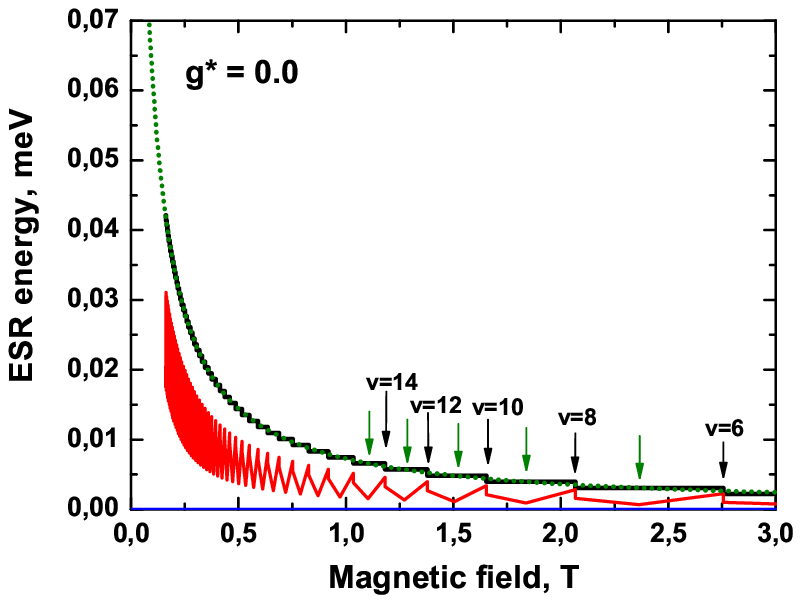}} % Here is how to import EPS art
  \end{minipage}
\caption{\label{Fig:3} (Color online) ESR energy as a function of magnetic field calculated for a model 2D system at different values of the g-factor: (a) -0.4, (b) 0.4, (c) -0.2, (d) 0.2, (e) 0.}
    \end{figure*}

In the opposite case ($n'$, $i'$)=($1$, $b$) and ($n$, $i$)=($0$, $a$), i.e., for $E_{0}^{(a)}>E_{1}^{(b)}$, Eq.~\eqref{eq:42} takes the form
\begin{gather}
\Delta^{(e-e)}_{SW}(0)=\int\limits_{0}^{+\infty}dxe^{-x^{2}/2}\tilde{h}_{1}^{(b)}\left(\dfrac{x}{a_B}\right)F\left(\dfrac{x}{a_B}\right)\times\notag\\ \times\Bigg\{\dfrac{x^2}{2}\left|\tilde{L}_{1,0}^{(b,a)}\left(\dfrac{x^2}{2}\right)\right|^2-\left|\tilde{L}_{1,1}^{(b,b)}\left(\dfrac{x^2}{2}\right)\right|^2+\tilde{L}_{1,1}^{(b,b)}\left(\dfrac{x^2}{2}\right)\Bigg\}.
\label{eq:44}\end{gather}
Taking into account Eqs.~\eqref{eq:24a} and \eqref{eq:24b}, one arrives at following form of $\Delta^{(e-e)}_{SW}(0)$:
\begin{gather}
\Delta^{(e-e)}_{SW}(0)=\int\limits_{0}^{+\infty}dxe^{-x^{2}/2}\tilde{h}_{1}^{(b)}\left(\dfrac{x}{a_B}\right)F\left(\dfrac{x}{a_B}\right)\times\notag\\ \times\dfrac{x^2}{2}\cos^2\varphi_1\left(2-\dfrac{x^2}{2}\cos^2\varphi_1\right)
\label{eq:45}\end{gather}
The earlier HFA results (see Eq.~\eqref{eq:37}) can be obtained by using Eq.~\eqref{eq:45} with $\tilde{h}_{1}^{(b)}(q)=-\nu_{1}^{(b)}$. We note that at $\nu_{1}^{(b)}=1$ this HFA expression becomes a Fourier component of the exact correlation function.

Analytical form for $\tilde{h}_{1}^{(b)}(q)$ at $\nu_{1}^{(b)}<1$ in the presence of SOI is unknown. However, the numerical many-body calculations, performed by Califano et al.\cite{q12} for \emph{negative} g-factor values, indicates that many-body correction to ESR energy  in the case of FQHE is \emph{positive}, in contrast to the HFA results $\Delta^{(e-e)}_{SW}(0)=-\nu_{1}^{(b)}\Delta^{(e-e)}_{b}<0$.

\section{\label{sec:RnD}Results and discussions}
To illustrate our theoretical results obtained in Sec.~\ref{sec:Theory}, we consider a 'model' 2D system with a zero thickness, in which $\epsilon=12.5$, $m^{*} = 0.067m_0$, $\alpha=0.005$ eV$\cdot${\AA} and 2DEG concentration is $4.0\cdot10^{11}$ cm$^{-2}$. The g-factor values are assumed to vary in the range from -0.4 to 0.4. These parameters are typical for 2D systems based on GaAs/AlGaAs\cite{q21,q53,q54,q55,q56}. Note that to describe the actual 2D systems one should take into account the nonzero-thickness effect. However, the model 2D system is quite sufficient for gaining a principle understanding of \emph{e-e} interaction effects in ESR.

\begin{figure}
\includegraphics [width=1.08\columnwidth, keepaspectratio] {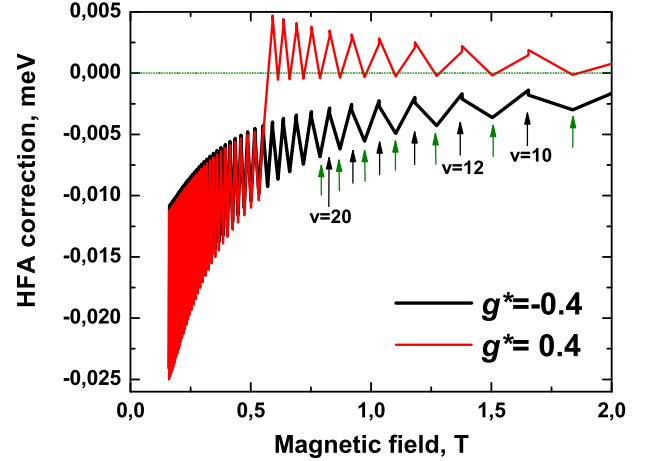}% Here is how to import EPS art
\caption{\label{Fig:4} (Color online) Oscillating behaviour of HFA corrections as a function of magnetic field at different values of g-factor.}
\end{figure}

\subsection{Electron spin resonance}
Figure~\ref{Fig:3} illustrates the ESR energy at different g-factor values as a function of magnetic field in the model 2D system. Black curve corresponds to the single-electron ESR energy, while the red one is the ESR energy calculated within HFA. Zeeman energy is marked by blue curve. The dotted green curve is described by Eq.~\eqref{eq:15}. The black and green arrows indicate the magnetic field values corresponding to the even and odd LL filling factors respectively.

As mentioned above, we restrict our consideration to the range of magnetic fields corresponding to $Y>Y_2(n_F)$, where $n_F$ is the number of the pair of spin-split LLs crossing the Fermi level. At a fixed value of the magnetic field, $n_F$ is found from the condition:
\begin{equation*}
2\pi a_B^2\sum_{n=0}^{n_F}\left(\nu_{n}^{(a)}+\nu_{n+1}^{(b)}\right)=n_S,
\end{equation*}
where $n_S$ is the 2DEG concentration. For the assigned parameters of the model 2D system and at $n_S=4.0\cdot10^{11}$ cm$^{-2}$ the value of dimensionless parameter $Y=Y_2(n_F)$ was reached at $n_F=51$ and $B\approx0.16$ T.

   \begin{figure*}
  \begin{minipage}{0.49\linewidth} %0.49\linewidth
\center{\Large{\textbf{(a)}} \includegraphics [width=0.95\columnwidth, keepaspectratio] {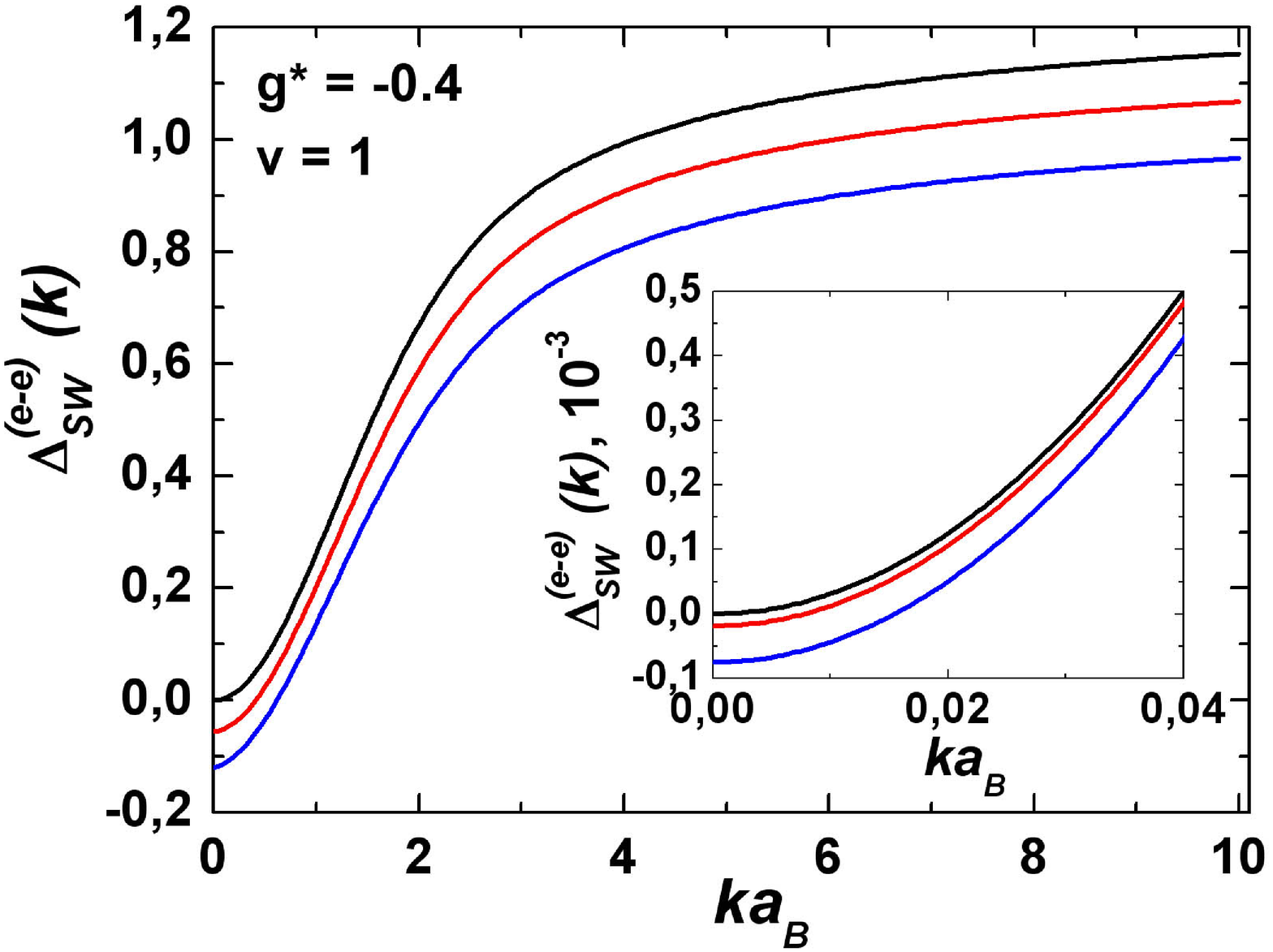}} % Here is how to import EPS art
   \end{minipage}
  \hfill
  \begin{minipage}{0.49\linewidth}
\center{\Large{\textbf{(b)}} \includegraphics [width=0.95\columnwidth, keepaspectratio] {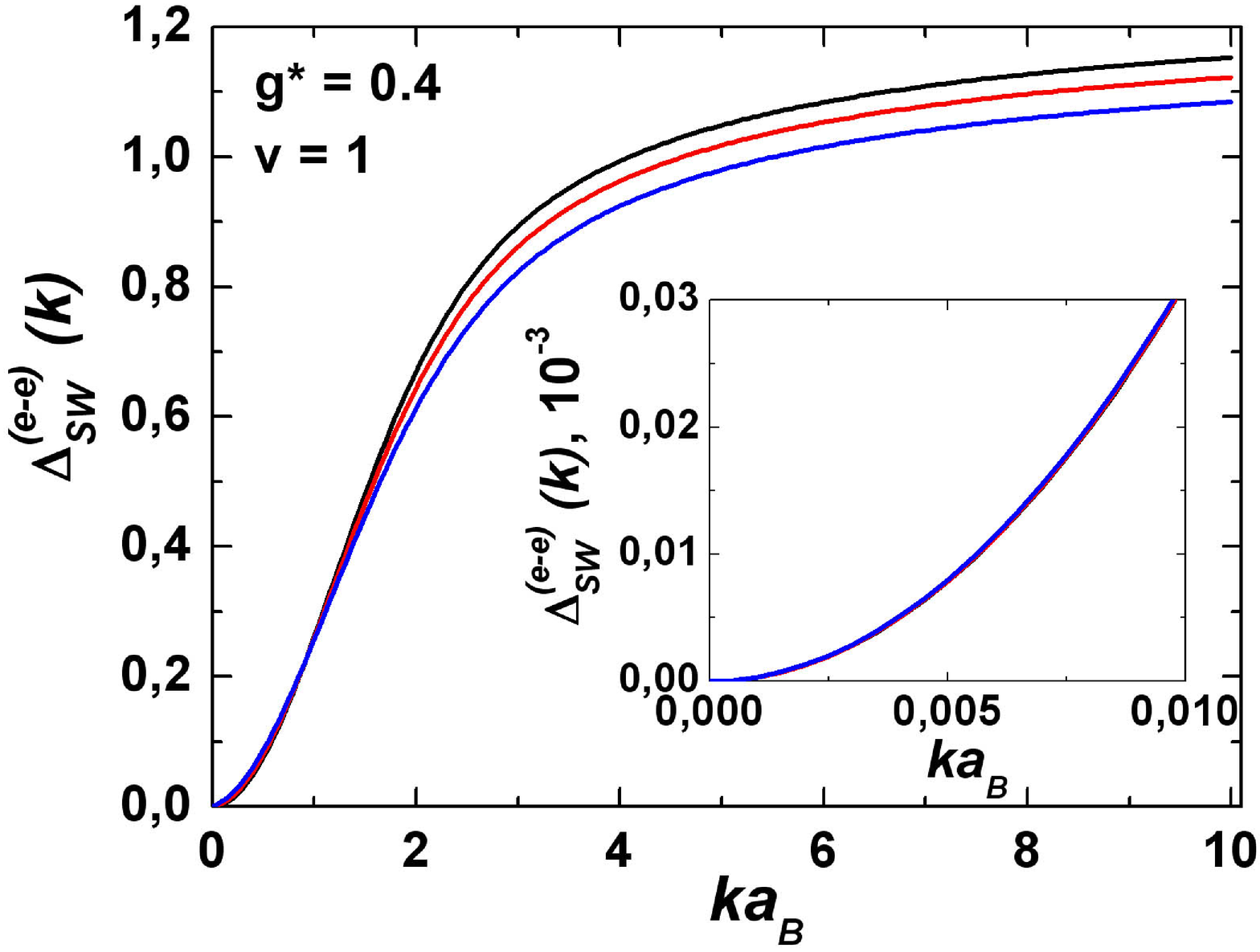}} % Here is how to import EPS art
  \end{minipage}
  \vfill
  \begin{minipage}{0.49\linewidth}
\center{\Large{\textbf{(c)}} \includegraphics [width=0.95\columnwidth, keepaspectratio] {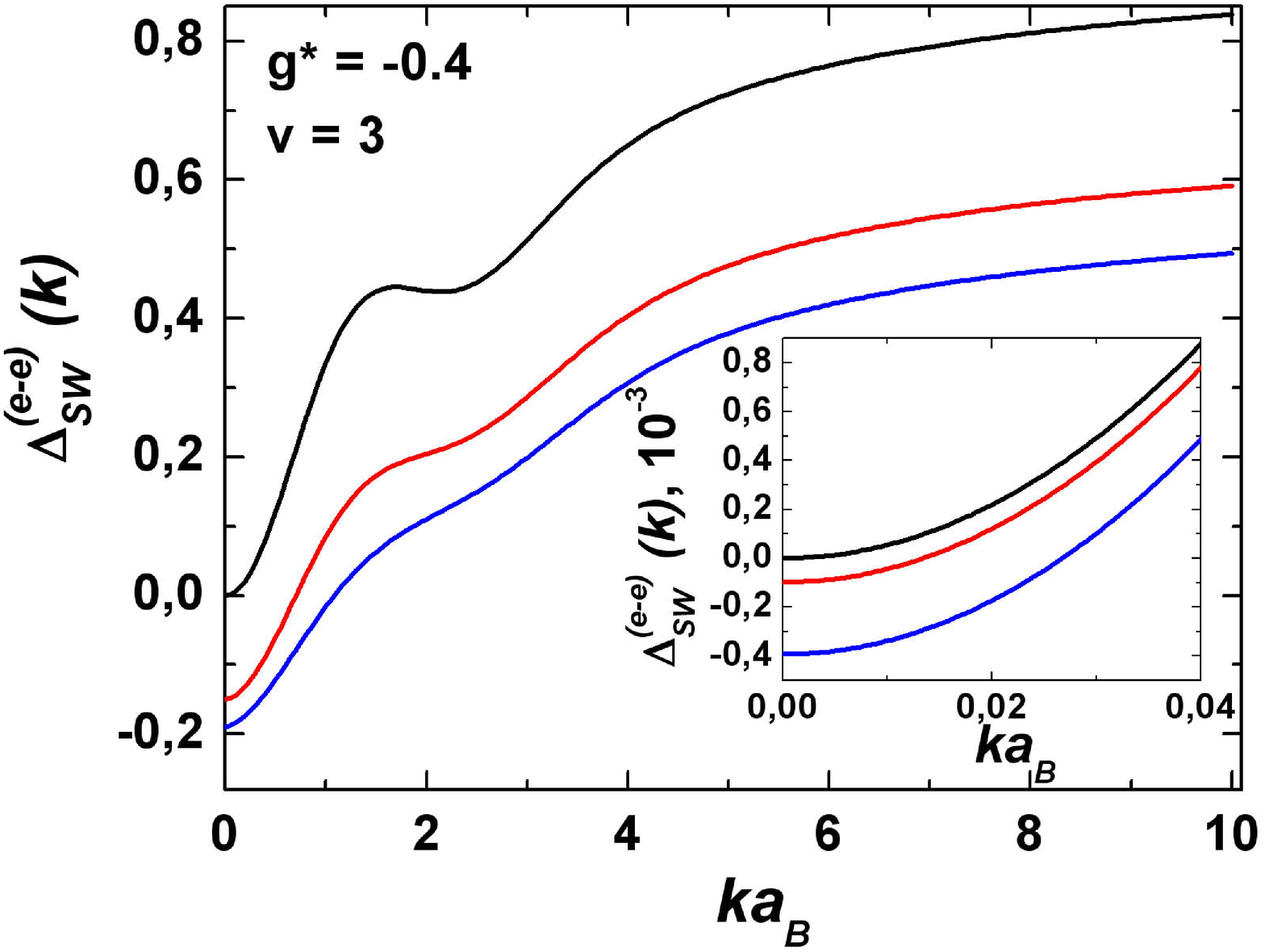}} % Here is how to import EPS art
  \end{minipage}
   \hfill
  \begin{minipage}{0.49\linewidth}
\center{\Large{\textbf{(d)}} \includegraphics [width=0.95\columnwidth, keepaspectratio] {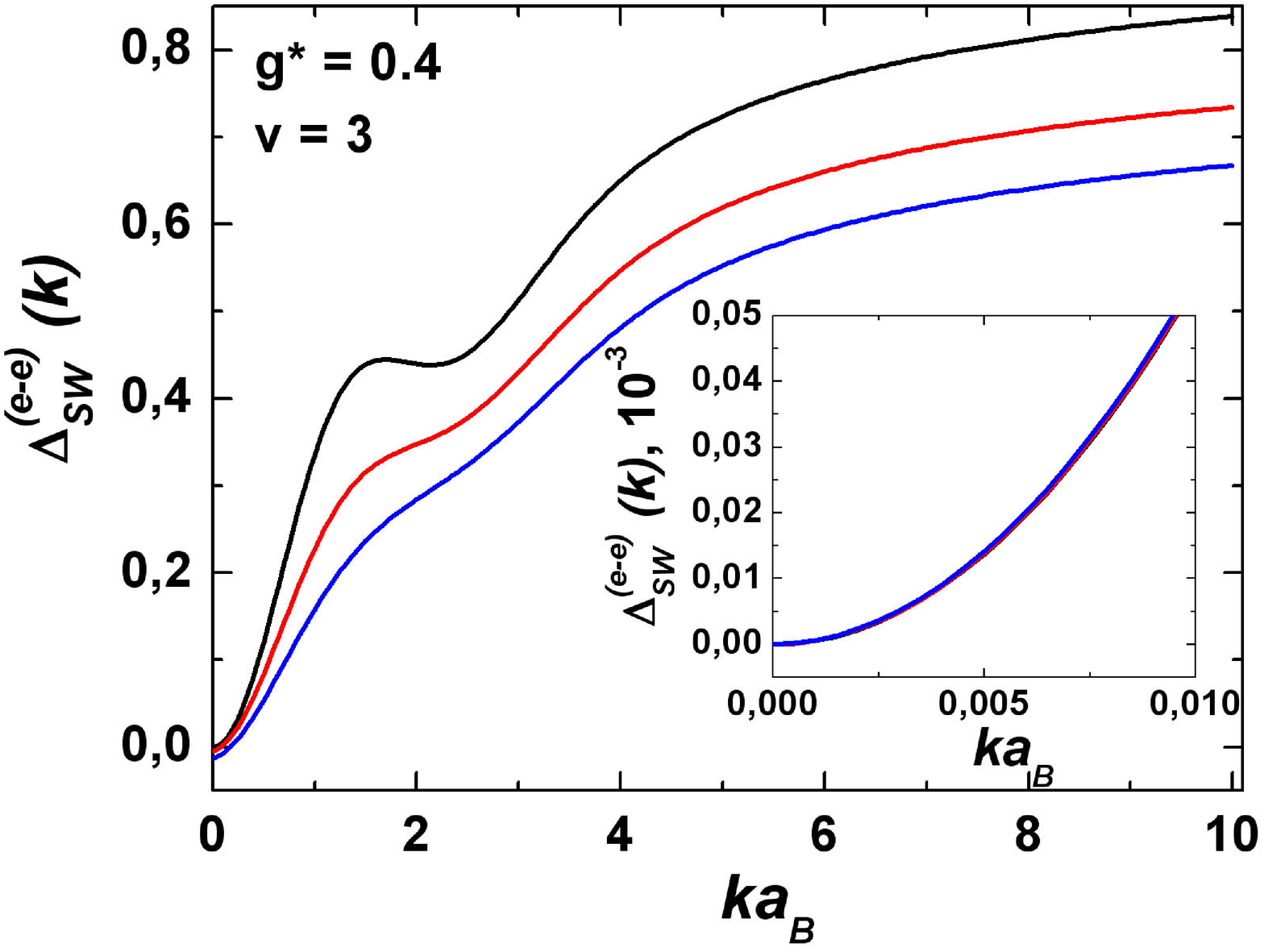}} % Here is how to import EPS art
  \end{minipage}
\caption{\label{Fig:5} (Color online) The energy of SW excitations $\Delta^{(e-e)}_{SW}(k)$ (in units of $e^2/\epsilon a_B$) counted from single-electron ESR energies in 2D system with BR spin splitting at odd-valued filling factors of the LLs ($\nu=$1 and 3) for different values of g-factor and BR constant: $\alpha=0$ eV$\cdot${\AA} (black curve), 0.3 eV$\cdot${\AA} (red curve), 0.5 eV$\cdot${\AA} (blue curve). The insets show the energies of SW excitation at small values of wave vector $k$ calculated for BR spin splitting values $\alpha=0$ (black curve), 0.005 (red curve), 0.01 (blue curve) eV$\cdot${\AA}, which are relevant for 2D systems based on GaAs/AlGaAs\cite{q21}.}
    \end{figure*}

As seen from Fig.~\ref{Fig:3}, the behavior of the single-electron ESR energy in a magnetic field is determined by the sign of the 2D electrons g-factor. At $g^{*}\leq 0$, as the magnetic field increases from zero, the single-electron ESR energy smoothly drops, tending to the value of Zeeman splitting in high magnetic fields. If the 2D electron g-factor is positive, the single-electron ESR energy is a non-smooth function of the magnetic field, exhibiting a 'V-shaped' behavior in the vicinity of critical magnetic field $B_{cr}$, where the single-electron ESR energy vanishes, its value obeying the condition $Y=Y_1(n_F)$.

The 'jump' features at the black curve at even-valued filling factors appear due to the dependence of the single-electron ESR energy on LL index n. As the magnetic field increases, the Fermi level 'jumps' from one pair of spin-split LLs to the lower-lying pair having a different value of spin splitting. This causes a sharp rise of the spin splitting energy at the Fermi level, followed by an abrupt change in the ESR energy. It should be noted that the weak dependence of the LL spin splitting on LL index $n$ at $\nu<6$ provides that the single-electron ESR energy in the model system is reasonably well described by Eq.~\eqref{eq:15} at any values of the g-factor.

Taking the \emph{e-e} interaction into account leads to substantial renormalization of the ESR energy in the region of magnetic fields, where the BR spin splitting has a noticeable effect on single-electron ESR energy. The oscillatory behavior of many-body correction $\Delta^{(e-e)}_{SW}(0)$ in a magnetic field is due to the oscillating difference in the filling factors of the LLs involved in the ESR transition $\nu_{n_F}^{(a)}-\nu_{n_F+1}^{(b)}$. At zero temperature in the absence of disorder in the 2D system, the densities of states at LLs ($n_F$,~$a$) and ($n_F+1$,~$b$) are the Dirac delta functions. Therefore, they never overlap, which, in particular, leads to existence of $\Delta^{(e-e)}_{SW}(0)$ oscillations even in an arbitrarily weak magnetic field. Once the disorder is taken into account, as is the case in the work by Antoniou and MacDonald\cite{q52}, the oscillations of $\nu_{n_F}^{(a)}-\nu_{n_F+1}^{(b)}$ and ESR energy are smeared out through overlapping of the density of states, which increases in weak magnetic fields. We also note that the disorder in a 2D system with BR splitting also causes a positive shift of the single-electron ESR energy\cite{q58}.

Many-body correction to the ESR energy, $\Delta^{(e-e)}_{SW}(0)$, strongly depends on a sign of the 2D electron g-factor. At $g^{*}\leq 0$ the \emph{e-e} interaction induces reduction of the ESR energy at all values of the magnetic field. We note that in 2D narrow-gap semiconductor systems with BR spin splitting the g-factor also has a negative sign, but the \emph{e-e} interaction in them leads to enhancement of the ESR energy due to additional effect of subband nonparabolicity\cite{q11}.

With a positive g-factor the contribution of \emph{e-e} interaction in the ESR energy can be both positive and negative, depending on a magnetic field. In the region of weak magnetic fields corresponding to $B\ll B_{cr}$, the \emph{e-e} interaction reduces the ESR energy as compared with the single-electron values. In high magnetic fields $B\gg B_{cr}$ the ESR energy is reduced only in the vicinity of the odd-valued LL filling factor. At other values of $\nu>1$ the \emph{e-e} interaction induces enhancement of the ESR energy.

In the vicinity of $B_{cr}$ the ESR energy has a complex behavior in the magnetic field. Figure~\ref{Fig:4} shows oscillating behaviour of the many-body corrections at different g-factor values. The black curve corresponds to the HFA correction calculated for $g^{*}=-0.4$. The black and green arrows in Fig.~\ref{Fig:4} indicate the magnetic field values corresponding to the even and odd LL filling factors, respectively. The red curve is the many-body-correction calculated for $g^{*}=0.4$. Note that in this case $B_{cr}\approx 0.55$ T (see Fig.~\ref{Fig:3}b). It is clear from Fig.~\ref{Fig:4} that in the vicinity of $B_{cr}$ the many-body correction at a positive g-factor undergoes an abrupt change from positive to negative value, as the magnetic field decreases. Since in actual ESR experiments the absolute value of the energy is measured, competing values of the single-electron ESR energy and many-body correction yield a complex oscillation picture, which is observed in Figs~\ref{Fig:3}b and \ref{Fig:3}d.

\subsection{Spin-wave excitations}
Figure~\ref{Fig:5} presents the spin-wave energies $\Delta^{(e-e)}_{SW}(k)$ (in units of $e^2/\epsilon a_B$) in a 2D system with BR spin splitting at odd-valued filling factors of the LLs ($\nu=$1 and 3) for different values of g-factor ($g^{*}=\pm0.4$) and BR constant. The energy was counted from the single-electron values of the ESR energy. Black, red and blue curves correspond to the calculation results obtained at $\alpha=0$, 0.3, 0.5 eV$\cdot${\AA}, respectively.

As seen from Fig.~\ref{Fig:5}, the BR spin splitting significantly modifies the dispersion of SW excitations for large values of $\alpha$. For example, the spin wave dispersions in the vicinity of $k=0$ at $\nu=1$ are determined by three terms: $\Delta^{(e-e)}_{ab}(k)$, quadratic in $k$, the linear term $\Delta^{(e-e)}(k)$ and $\Delta^{(e-e)}_{b}$, which does not depend on the wave vector. The $\Delta^{(e-e)}(k)$ and $\Delta^{(e-e)}_{b}$ related contributions to the SW excitation energy increase with a rising $\alpha$. However, the contribution of $\Delta^{(e-e)}_{ab}(k)$ diminishes, which leads to distortion of the dispersion shape of the SW excitation.

In all cases of interest the spin-wave energy has a negative shift as compared with the values at $\alpha=0$. Note that according to Eq.~\eqref{eq:34b}, the shift tends to zero in the limit $k\rightarrow 0$ for $\nu=1$ and $g^{*}>0$. It is clear that for the same absolute values of g-factor and at a fixed BR constant the spin-wave energies are shifted to a greater extent at $g^{*}<0$ than is the case at positive g-factors. Especially this is explicitly seen in the range of $ka_B\ll 1$.

\begin{figure}
\includegraphics [width=0.95\columnwidth, keepaspectratio] {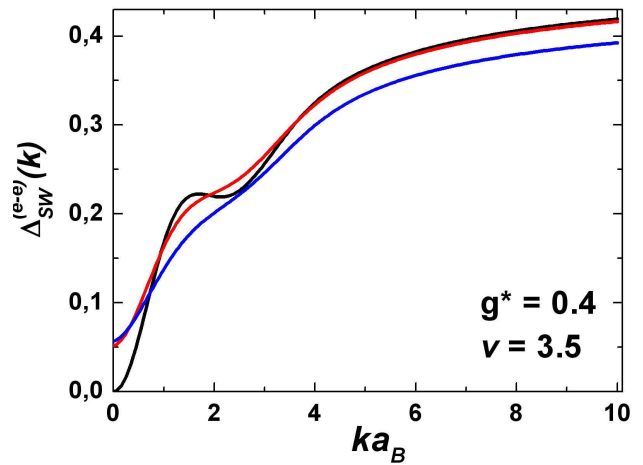}% Here is how to import EPS art
\caption{\label{Fig:6} (Color online) The energy of SW excitations $\Delta^{(e-e)}_{SW}(k)$ (in units of $e^2/\epsilon a_B$) counted from single-electron ESR energies in 2D system with $g^{*}=0.4$ at $\nu=3.5$ for different values of BR constant: $\alpha=0$ eV$\cdot${\AA} (black curve), 0.3 eV$\cdot${\AA} (red curve), 0.5 eV$\cdot${\AA} (blue curve).}
\end{figure}

We have considered the SW excitations at odd LL filling factors. As the LL filling factor deviates from the values of $\nu=1$ and 3 in a 2D system with $g^{*}<0$, the SW excitations are qualitatively described by the dispersion curves similar to those given in Fig.~\ref{Fig:5}(a, c). In 2D systems with positive g-factor values the dispersion curves of SW excitations are significantly modified by deviations from the odd-valued LL filling factor. As clearly seen from Fig.~\ref{Fig:6}, the dispersions have a positive shift at small values of the wave vector. In the limit of infinitely large wave vectors the shift is negative. With a growing value of the LL filling factor, the BR spin splitting influence on the dispersion curves of SW excitations increases.

\section{\label{sec:Summary}Conclusions}
We have studied a contribution of \emph{e-e} interaction to ESR energy and spin-wave excitations in perpendicular magnetic field in 2D systems with Bychkov-Rashba spin splitting. By using HFA, the many-body corrections to the single-particle ESR energy are found to be nonzero. We have found out that \emph{e-e} interaction in 2D systems with SIA can not only enhance the ESR energy, as it has been predicted for narrow-gap QWs\cite{q11}, but also can lead to the ESR energy reduction. We demonstrate that many-body ESR energy is an oscillating function of the magnetic field, which behavior is rather sensitive to the sign of g-factor and the LL filling factor. In particular, we have shown that \emph{e-e} interaction does not affect the ESR energy in the case of the lowest LL filling in 2D systems with positive g-factors. We derive analytical expressions for the case $\nu\leq 2$ and show that significant dependence of the dispersion curve of SW excitation on BR constant and g-factor values, especially for large values of $\alpha$.

The results obtained indicate that \emph{e-e} interaction in any case should be taken into consideration in analyzing of experiments on ESR in the regions of magnetic field, in which the BR constant $\alpha$ could be extracted from experimental data (cf. Ref.~\onlinecite{q59}).

\begin{acknowledgments}
The author is grateful to Vladimir Gavrilenko and Maksim Zholudev from the Institute for Physics of Microstructures RAS (Nizhny Novgorod), as well as Revaz Ramazashvili from Laboratoire de Physique Th\'{e}orique (Toulouse) for scientific discussions and valuable suggestions. This work is partially supported by the Russian Foundation for Basic Research (Grants 11-02-93111, 12-02-00940, 13-02-00894), by the Russian Academy of Sciences and by Russian Ministry of Education and Science (Grant HIII-4756.2012.2).
\end{acknowledgments}


\begin{thebibliography}{64}
\expandafter\ifx\csname natexlab\endcsname\relax\def\natexlab#1{#1}\fi
\expandafter\ifx\csname bibnamefont\endcsname\relax
  \def\bibnamefont#1{#1}\fi
\expandafter\ifx\csname bibfnamefont\endcsname\relax
  \def\bibfnamefont#1{#1}\fi
\expandafter\ifx\csname citenamefont\endcsname\relax
  \def\citenamefont#1{#1}\fi
\expandafter\ifx\csname url\endcsname\relax
  \def\url#1{\texttt{#1}}\fi
\expandafter\ifx\csname urlprefix\endcsname\relax\def\urlprefix{URL }\fi
\providecommand{\bibinfo}[2]{#2}
\providecommand{\eprint}[2][]{\url{#2}}

\bibitem[{\citenamefont{Bloch}(1930)}]{q1}
\bibinfo{author}{\bibfnamefont{F.}~\bibnamefont{Bloch}}, \bibinfo{journal}{Z.
  Phys.} \textbf{\bibinfo{volume}{61}}, \bibinfo{pages}{206}
  (\bibinfo{year}{1930}).

\bibitem[{\citenamefont{Longo and Kallin}(1993)}]{q2}
\bibinfo{author}{\bibfnamefont{J.~P.} \bibnamefont{Longo}} \bibnamefont{and}
  \bibinfo{author}{\bibfnamefont{C.}~\bibnamefont{Kallin}},
  \bibinfo{journal}{Phys. Rev. B} \textbf{\bibinfo{volume}{47}},
  \bibinfo{pages}{4429} (\bibinfo{year}{1993}).

\bibitem[{\citenamefont{Schliemann et~al.}(2003)\citenamefont{Schliemann,
  Egues, and Loss}}]{q3}
\bibinfo{author}{\bibfnamefont{J.}~\bibnamefont{Schliemann}},
  \bibinfo{author}{\bibfnamefont{J.~C.} \bibnamefont{Egues}}, \bibnamefont{and}
  \bibinfo{author}{\bibfnamefont{D.}~\bibnamefont{Loss}},
  \bibinfo{journal}{Phys. Rev. B} \textbf{\bibinfo{volume}{67}},
  \bibinfo{pages}{085302} (\bibinfo{year}{2003}).

\bibitem[{\citenamefont{Krishtopenko et~al.}(2011)\citenamefont{Krishtopenko,
  Gavrilenko, and Goiran}}]{q4}
\bibinfo{author}{\bibfnamefont{S.~S.} \bibnamefont{Krishtopenko}},
  \bibinfo{author}{\bibfnamefont{V.~I.} \bibnamefont{Gavrilenko}},
  \bibnamefont{and} \bibinfo{author}{\bibfnamefont{M.}~\bibnamefont{Goiran}},
  \bibinfo{journal}{J. Phys.: Condens. Matter} \textbf{\bibinfo{volume}{23}},
  \bibinfo{pages}{385601} (\bibinfo{year}{2011}).

\bibitem[{\citenamefont{Krishtopenko
  et~al.}(2012{\natexlab{a}})\citenamefont{Krishtopenko, Gavrilenko, and
  Goiran}}]{q5}
\bibinfo{author}{\bibfnamefont{S.~S.} \bibnamefont{Krishtopenko}},
  \bibinfo{author}{\bibfnamefont{V.~I.} \bibnamefont{Gavrilenko}},
  \bibnamefont{and} \bibinfo{author}{\bibfnamefont{M.}~\bibnamefont{Goiran}},
  \bibinfo{journal}{J. Phys.: Condens. Matter} \textbf{\bibinfo{volume}{24}},
  \bibinfo{pages}{135601} (\bibinfo{year}{2012}{\natexlab{a}}).

\bibitem[{\citenamefont{Krishtopenko
  et~al.}(2012{\natexlab{b}})\citenamefont{Krishtopenko, Kalinin, Gavrilenko,
  Sadofyev, and Goiran}}]{q6}
\bibinfo{author}{\bibfnamefont{S.~S.} \bibnamefont{Krishtopenko}},
  \bibinfo{author}{\bibfnamefont{K.~P.} \bibnamefont{Kalinin}},
  \bibinfo{author}{\bibfnamefont{V.~I.} \bibnamefont{Gavrilenko}},
  \bibinfo{author}{\bibfnamefont{Y.~G.} \bibnamefont{Sadofyev}},
  \bibnamefont{and} \bibinfo{author}{\bibfnamefont{M.}~\bibnamefont{Goiran}},
  \bibinfo{journal}{Semiconductors} \textbf{\bibinfo{volume}{46}},
  \bibinfo{pages}{1163} (\bibinfo{year}{2012}{\natexlab{b}}).

\bibitem[{\citenamefont{Aleshkin et~al.}(2008)\citenamefont{Aleshkin,
  Gavrilenko, Ikonnikov, Krishtopenko, Sadofyev, and Spirin}}]{q7}
\bibinfo{author}{\bibfnamefont{V.~Y.} \bibnamefont{Aleshkin}},
  \bibinfo{author}{\bibfnamefont{V.~I.} \bibnamefont{Gavrilenko}},
  \bibinfo{author}{\bibfnamefont{A.~V.} \bibnamefont{Ikonnikov}},
  \bibinfo{author}{\bibfnamefont{S.~S.} \bibnamefont{Krishtopenko}},
  \bibinfo{author}{\bibfnamefont{Y.~G.} \bibnamefont{Sadofyev}},
  \bibnamefont{and} \bibinfo{author}{\bibfnamefont{K.~E.}
  \bibnamefont{Spirin}}, \bibinfo{journal}{Semiconductors}
  \textbf{\bibinfo{volume}{42}}, \bibinfo{pages}{828} (\bibinfo{year}{2008}).

\bibitem[{\citenamefont{Gavrilenko et~al.}(2011)\citenamefont{Gavrilenko,
  Krishtopenko, and Goiran}}]{q8}
\bibinfo{author}{\bibfnamefont{V.~I.} \bibnamefont{Gavrilenko}},
  \bibinfo{author}{\bibfnamefont{S.~S.} \bibnamefont{Krishtopenko}},
  \bibnamefont{and} \bibinfo{author}{\bibfnamefont{M.}~\bibnamefont{Goiran}},
  \bibinfo{journal}{Semiconductors} \textbf{\bibinfo{volume}{45}},
  \bibinfo{pages}{110} (\bibinfo{year}{2011}).

\bibitem[{\citenamefont{Krishtopenko
  et~al.}(2012{\natexlab{c}})\citenamefont{Krishtopenko, Gavrilenko, and
  Goiran}}]{q9}
\bibinfo{author}{\bibfnamefont{S.~S.} \bibnamefont{Krishtopenko}},
  \bibinfo{author}{\bibfnamefont{V.~I.} \bibnamefont{Gavrilenko}},
  \bibnamefont{and} \bibinfo{author}{\bibfnamefont{M.}~\bibnamefont{Goiran}},
  \bibinfo{journal}{Solid State Phenomena} \textbf{\bibinfo{volume}{190}},
  \bibinfo{pages}{554} (\bibinfo{year}{2012}{\natexlab{c}}).

\bibitem[{\citenamefont{Krishtopenko
  et~al.}(2012{\natexlab{d}})\citenamefont{Krishtopenko, Gavrilenko, and
  Goiran}}]{q10}
\bibinfo{author}{\bibfnamefont{S.~S.} \bibnamefont{Krishtopenko}},
  \bibinfo{author}{\bibfnamefont{V.~I.} \bibnamefont{Gavrilenko}},
  \bibnamefont{and} \bibinfo{author}{\bibfnamefont{M.}~\bibnamefont{Goiran}},
  \bibinfo{journal}{J. Phys.: Condens. Matter} \textbf{\bibinfo{volume}{24}},
  \bibinfo{pages}{252201} (\bibinfo{year}{2012}{\natexlab{d}}).

\bibitem[{\citenamefont{Krishtopenko et~al.}(2013)\citenamefont{Krishtopenko,
  Gavrilenko, and Goiran}}]{q11}
\bibinfo{author}{\bibfnamefont{S.~S.} \bibnamefont{Krishtopenko}},
  \bibinfo{author}{\bibfnamefont{V.~I.} \bibnamefont{Gavrilenko}},
  \bibnamefont{and} \bibinfo{author}{\bibfnamefont{M.}~\bibnamefont{Goiran}},
  \bibinfo{journal}{Phys. Rev. B} \textbf{\bibinfo{volume}{87}},
  \bibinfo{pages}{155113} (\bibinfo{year}{2013}).

\bibitem[{\citenamefont{Califano et~al.}(2005)\citenamefont{Califano,
  Chakraborty, and Pietil\"ainen}}]{q12}
\bibinfo{author}{\bibfnamefont{M.}~\bibnamefont{Califano}},
  \bibinfo{author}{\bibfnamefont{T.}~\bibnamefont{Chakraborty}},
  \bibnamefont{and}
  \bibinfo{author}{\bibfnamefont{P.}~\bibnamefont{Pietil\"ainen}},
  \bibinfo{journal}{Phys. Rev. Lett.} \textbf{\bibinfo{volume}{94}},
  \bibinfo{pages}{246801} (\bibinfo{year}{2005}).

\bibitem[{\citenamefont{Chesi and Loss}(2008)}]{q13}
\bibinfo{author}{\bibfnamefont{S.}~\bibnamefont{Chesi}} \bibnamefont{and}
  \bibinfo{author}{\bibfnamefont{D.}~\bibnamefont{Loss}},
  \bibinfo{journal}{Phys. Rev. Lett.} \textbf{\bibinfo{volume}{101}},
  \bibinfo{pages}{146803} (\bibinfo{year}{2008}).

\bibitem[{\citenamefont{Ito et~al.}(2010)\citenamefont{Ito, Nomura, and
  Shibata}}]{q14}
\bibinfo{author}{\bibfnamefont{T.}~\bibnamefont{Ito}},
  \bibinfo{author}{\bibfnamefont{K.}~\bibnamefont{Nomura}}, \bibnamefont{and}
  \bibinfo{author}{\bibfnamefont{N.}~\bibnamefont{Shibata}},
  \bibinfo{journal}{J. Phys. Soc. Jpn.} \textbf{\bibinfo{volume}{79}},
  \bibinfo{pages}{073003} (\bibinfo{year}{2010}).

\bibitem[{\citenamefont{Ito et~al.}(2012)\citenamefont{Ito, Nomura, and
  Shibata}}]{q15}
\bibinfo{author}{\bibfnamefont{T.}~\bibnamefont{Ito}},
  \bibinfo{author}{\bibfnamefont{K.}~\bibnamefont{Nomura}}, \bibnamefont{and}
  \bibinfo{author}{\bibfnamefont{N.}~\bibnamefont{Shibata}},
  \bibinfo{journal}{J. Phys. Soc. Jpn.} \textbf{\bibinfo{volume}{81}},
  \bibinfo{pages}{034713} (\bibinfo{year}{2012}).

\bibitem[{\citenamefont{Califano et~al.}(2006)\citenamefont{Califano,
  Chakraborty, Pietil\"ainen, and Hu}}]{q16}
\bibinfo{author}{\bibfnamefont{M.}~\bibnamefont{Califano}},
  \bibinfo{author}{\bibfnamefont{T.}~\bibnamefont{Chakraborty}},
  \bibinfo{author}{\bibfnamefont{P.}~\bibnamefont{Pietil\"ainen}},
  \bibnamefont{and} \bibinfo{author}{\bibfnamefont{C.-M.} \bibnamefont{Hu}},
  \bibinfo{journal}{Phys. Rev. B} \textbf{\bibinfo{volume}{73}},
  \bibinfo{pages}{113315} (\bibinfo{year}{2006}).

\bibitem[{\citenamefont{Krishtopenko}(2013{\natexlab{a}})}]{q17}
\bibinfo{author}{\bibfnamefont{S.~S.} \bibnamefont{Krishtopenko}},
  \bibinfo{journal}{J. Phys.: Condens. Matter} \textbf{\bibinfo{volume}{25}},
  \bibinfo{pages}{105601} (\bibinfo{year}{2013}{\natexlab{a}}).

\bibitem[{\citenamefont{Bychkov and Rashba}(1984)}]{q18}
\bibinfo{author}{\bibfnamefont{Y.~A.} \bibnamefont{Bychkov}} \bibnamefont{and}
  \bibinfo{author}{\bibfnamefont{E.~I.} \bibnamefont{Rashba}},
  \bibinfo{journal}{J. Phys. C: Solid State Phys.}
  \textbf{\bibinfo{volume}{17}}, \bibinfo{pages}{6039} (\bibinfo{year}{1984}).

\bibitem[{\citenamefont{Nitta et~al.}(1997)\citenamefont{Nitta, Akazaki,
  Takayanagi, and Enoki}}]{q19}
\bibinfo{author}{\bibfnamefont{J.}~\bibnamefont{Nitta}},
  \bibinfo{author}{\bibfnamefont{T.}~\bibnamefont{Akazaki}},
  \bibinfo{author}{\bibfnamefont{H.}~\bibnamefont{Takayanagi}},
  \bibnamefont{and} \bibinfo{author}{\bibfnamefont{T.}~\bibnamefont{Enoki}},
  \bibinfo{journal}{Phys. Rev. Lett.} \textbf{\bibinfo{volume}{78}},
  \bibinfo{pages}{1335} (\bibinfo{year}{1997}).

\bibitem[{\citenamefont{Papadakis et~al.}(1999)\citenamefont{Papadakis,
  De~Poortere, Manoharan, Shayegan, and Winkler}}]{q20}
\bibinfo{author}{\bibfnamefont{S.~J.} \bibnamefont{Papadakis}},
  \bibinfo{author}{\bibfnamefont{E.~P.} \bibnamefont{De~Poortere}},
  \bibinfo{author}{\bibfnamefont{H.~C.} \bibnamefont{Manoharan}},
  \bibinfo{author}{\bibfnamefont{M.}~\bibnamefont{Shayegan}}, \bibnamefont{and}
  \bibinfo{author}{\bibfnamefont{R.}~\bibnamefont{Winkler}},
  \bibinfo{journal}{Science} \textbf{\bibinfo{volume}{283}},
  \bibinfo{pages}{2056} (\bibinfo{year}{1999}).

\bibitem[{\citenamefont{Studer et~al.}(2009{\natexlab{a}})\citenamefont{Studer,
  Salis, Ensslin, Driscoll, and Gossard}}]{q21}
\bibinfo{author}{\bibfnamefont{M.}~\bibnamefont{Studer}},
  \bibinfo{author}{\bibfnamefont{G.}~\bibnamefont{Salis}},
  \bibinfo{author}{\bibfnamefont{K.}~\bibnamefont{Ensslin}},
  \bibinfo{author}{\bibfnamefont{D.~C.} \bibnamefont{Driscoll}},
  \bibnamefont{and} \bibinfo{author}{\bibfnamefont{A.~C.}
  \bibnamefont{Gossard}}, \bibinfo{journal}{Phys. Rev. Lett.}
  \textbf{\bibinfo{volume}{103}}, \bibinfo{pages}{027201}
  (\bibinfo{year}{2009}{\natexlab{a}}).

\bibitem[{\citenamefont{Tuttle et~al.}(1989)\citenamefont{Tuttle, Kroemer, and
  English}}]{q22}
\bibinfo{author}{\bibfnamefont{G.}~\bibnamefont{Tuttle}},
  \bibinfo{author}{\bibfnamefont{H.}~\bibnamefont{Kroemer}}, \bibnamefont{and}
  \bibinfo{author}{\bibfnamefont{J.~H.} \bibnamefont{English}},
  \bibinfo{journal}{J. Appl. Phys.} \textbf{\bibinfo{volume}{65}},
  \bibinfo{pages}{5239} (\bibinfo{year}{1989}).

\bibitem[{\citenamefont{Gavrilenko et~al.}(2010)\citenamefont{Gavrilenko,
  Ikonnikov, Krishtopenko, Lastovkin, Marem'yanin, Sadofyev, and Spirin}}]{q23}
\bibinfo{author}{\bibfnamefont{V.~I.} \bibnamefont{Gavrilenko}},
  \bibinfo{author}{\bibfnamefont{A.~V.} \bibnamefont{Ikonnikov}},
  \bibinfo{author}{\bibfnamefont{S.~S.} \bibnamefont{Krishtopenko}},
  \bibinfo{author}{\bibfnamefont{A.~A.} \bibnamefont{Lastovkin}},
  \bibinfo{author}{\bibfnamefont{K.~V.} \bibnamefont{Marem'yanin}},
  \bibinfo{author}{\bibfnamefont{Y.~G.} \bibnamefont{Sadofyev}},
  \bibnamefont{and} \bibinfo{author}{\bibfnamefont{K.~E.}
  \bibnamefont{Spirin}}, \bibinfo{journal}{Semiconductors}
  \textbf{\bibinfo{volume}{44}}, \bibinfo{pages}{616} (\bibinfo{year}{2010}).

\bibitem[{\citenamefont{Spirin et~al.}(2012)\citenamefont{Spirin, Kalinin,
  Krishtopenko, Maremyanin, Gavrilenko, and Sadofyev}}]{q24}
\bibinfo{author}{\bibfnamefont{K.~E.} \bibnamefont{Spirin}},
  \bibinfo{author}{\bibfnamefont{K.~P.} \bibnamefont{Kalinin}},
  \bibinfo{author}{\bibfnamefont{S.~S.} \bibnamefont{Krishtopenko}},
  \bibinfo{author}{\bibfnamefont{K.~V.} \bibnamefont{Maremyanin}},
  \bibinfo{author}{\bibfnamefont{V.~I.} \bibnamefont{Gavrilenko}},
  \bibnamefont{and} \bibinfo{author}{\bibfnamefont{Y.~G.}
  \bibnamefont{Sadofyev}}, \bibinfo{journal}{Semiconductors}
  \textbf{\bibinfo{volume}{46}}, \bibinfo{pages}{1396} (\bibinfo{year}{2012}).

\bibitem[{\citenamefont{Pfeffer and Zawadzki}(2003)}]{q25}
\bibinfo{author}{\bibfnamefont{P.}~\bibnamefont{Pfeffer}} \bibnamefont{and}
  \bibinfo{author}{\bibfnamefont{W.}~\bibnamefont{Zawadzki}},
  \bibinfo{journal}{Phys. Rev. B} \textbf{\bibinfo{volume}{68}},
  \bibinfo{pages}{035315} (\bibinfo{year}{2003}).

\bibitem[{\citenamefont{Zawadzki and Pfeffer}(2004)}]{q26}
\bibinfo{author}{\bibfnamefont{W.}~\bibnamefont{Zawadzki}} \bibnamefont{and}
  \bibinfo{author}{\bibfnamefont{P.}~\bibnamefont{Pfeffer}},
  \bibinfo{journal}{Semicond. Sci. Technol.} \textbf{\bibinfo{volume}{19}},
  \bibinfo{pages}{R1} (\bibinfo{year}{2004}).

\bibitem[{\citenamefont{Krishtopenko}(2013{\natexlab{b}})}]{q60}
\bibinfo{author}{\bibfnamefont{S.~S.} \bibnamefont{Krishtopenko}},
  \bibinfo{journal}{J. Phys.: Condens. Matter} \textbf{\bibinfo{volume}{25}},
  \bibinfo{pages}{365602} (\bibinfo{year}{2013}{\natexlab{b}}).

\bibitem[{\citenamefont{Winkler}(1996)}]{q27}
\bibinfo{author}{\bibfnamefont{R.}~\bibnamefont{Winkler}},
  \bibinfo{journal}{Surface Science} \textbf{\bibinfo{volume}{361/362}},
  \bibinfo{pages}{411 } (\bibinfo{year}{1996}).

\bibitem[{\citenamefont{Winkler}(Springer, Berlin, Heidelberg, 2003)}]{q28}
\bibinfo{author}{\bibfnamefont{R.}~\bibnamefont{Winkler}},
  \bibinfo{journal}{\emph{Spin Orbit Coupling Effects in Two-Dimensional
  Electron and Hole Systems}}  (\bibinfo{year}{Springer, Berlin, Heidelberg,
  2003}).

\bibitem[{\citenamefont{Yang et~al.}(1993)\citenamefont{Yang, Lin-Chung,
  Shanabrook, Waterman, Wagner, and Moore}}]{q29}
\bibinfo{author}{\bibfnamefont{M.~J.} \bibnamefont{Yang}},
  \bibinfo{author}{\bibfnamefont{P.~J.} \bibnamefont{Lin-Chung}},
  \bibinfo{author}{\bibfnamefont{B.~V.} \bibnamefont{Shanabrook}},
  \bibinfo{author}{\bibfnamefont{J.~R.} \bibnamefont{Waterman}},
  \bibinfo{author}{\bibfnamefont{R.~J.} \bibnamefont{Wagner}},
  \bibnamefont{and} \bibinfo{author}{\bibfnamefont{W.~J.} \bibnamefont{Moore}},
  \bibinfo{journal}{Phys. Rev. B} \textbf{\bibinfo{volume}{47}},
  \bibinfo{pages}{1691} (\bibinfo{year}{1993}).

\bibitem[{\citenamefont{Ikonnikov et~al.}(2010)\citenamefont{Ikonnikov,
  Krishtopenko, Gavrilenko, Sadofyev, Vasilyev, Orlita, and Knap}}]{q30}
\bibinfo{author}{\bibfnamefont{A.~V.} \bibnamefont{Ikonnikov}},
  \bibinfo{author}{\bibfnamefont{S.~S.} \bibnamefont{Krishtopenko}},
  \bibinfo{author}{\bibfnamefont{V.~I.} \bibnamefont{Gavrilenko}},
  \bibinfo{author}{\bibfnamefont{Y.~G.} \bibnamefont{Sadofyev}},
  \bibinfo{author}{\bibfnamefont{Y.~B.} \bibnamefont{Vasilyev}},
  \bibinfo{author}{\bibfnamefont{M.}~\bibnamefont{Orlita}}, \bibnamefont{and}
  \bibinfo{author}{\bibfnamefont{W.}~\bibnamefont{Knap}}, \bibinfo{journal}{J.
  Low Temp. Phys.} \textbf{\bibinfo{volume}{159}}, \bibinfo{pages}{197}
  (\bibinfo{year}{2010}).

\bibitem[{\citenamefont{Krishtopenko
  et~al.}(2012{\natexlab{e}})\citenamefont{Krishtopenko, Ikonnikov, Maremyanin,
  Spirin, Gavrilenko, Sadofyev, Goiran, Sadowsky, and Vasilyev}}]{q31}
\bibinfo{author}{\bibfnamefont{S.~S.} \bibnamefont{Krishtopenko}},
  \bibinfo{author}{\bibfnamefont{A.~V.} \bibnamefont{Ikonnikov}},
  \bibinfo{author}{\bibfnamefont{A.~V.} \bibnamefont{Maremyanin}},
  \bibinfo{author}{\bibfnamefont{K.~E.} \bibnamefont{Spirin}},
  \bibinfo{author}{\bibfnamefont{V.~I.} \bibnamefont{Gavrilenko}},
  \bibinfo{author}{\bibfnamefont{Y.~G.} \bibnamefont{Sadofyev}},
  \bibinfo{author}{\bibfnamefont{M.}~\bibnamefont{Goiran}},
  \bibinfo{author}{\bibfnamefont{M.}~\bibnamefont{Sadowsky}}, \bibnamefont{and}
  \bibinfo{author}{\bibfnamefont{Y.~B.} \bibnamefont{Vasilyev}},
  \bibinfo{journal}{J. Appl. Phys.} \textbf{\bibinfo{volume}{111}},
  \bibinfo{eid}{093711} (\bibinfo{year}{2012}{\natexlab{e}}).

\bibitem[{\citenamefont{Eremeev et~al.}(2012)\citenamefont{Eremeev, Nechaev,
  Koroteev, Echenique, and Chulkov}}]{q32}
\bibinfo{author}{\bibfnamefont{S.~V.} \bibnamefont{Eremeev}},
  \bibinfo{author}{\bibfnamefont{I.~A.} \bibnamefont{Nechaev}},
  \bibinfo{author}{\bibfnamefont{Y.~M.} \bibnamefont{Koroteev}},
  \bibinfo{author}{\bibfnamefont{P.~M.} \bibnamefont{Echenique}},
  \bibnamefont{and} \bibinfo{author}{\bibfnamefont{E.~V.}
  \bibnamefont{Chulkov}}, \bibinfo{journal}{Phys. Rev. Lett.}
  \textbf{\bibinfo{volume}{108}}, \bibinfo{pages}{246802}
  (\bibinfo{year}{2012}).

\bibitem[{\citenamefont{Crepaldi et~al.}(2012)\citenamefont{Crepaldi,
  Moreschini, Aut\`es, Tournier-Colletta, Moser, Virk, Berger, Bugnon, Chang,
  Kern et~al.}}]{q33}
\bibinfo{author}{\bibfnamefont{A.}~\bibnamefont{Crepaldi}},
  \bibinfo{author}{\bibfnamefont{L.}~\bibnamefont{Moreschini}},
  \bibinfo{author}{\bibfnamefont{G.}~\bibnamefont{Aut\`es}},
  \bibinfo{author}{\bibfnamefont{C.}~\bibnamefont{Tournier-Colletta}},
  \bibinfo{author}{\bibfnamefont{S.}~\bibnamefont{Moser}},
  \bibinfo{author}{\bibfnamefont{N.}~\bibnamefont{Virk}},
  \bibinfo{author}{\bibfnamefont{H.}~\bibnamefont{Berger}},
  \bibinfo{author}{\bibfnamefont{P.}~\bibnamefont{Bugnon}},
  \bibinfo{author}{\bibfnamefont{Y.~J.} \bibnamefont{Chang}},
  \bibinfo{author}{\bibfnamefont{K.}~\bibnamefont{Kern}}, \bibnamefont{et~al.},
  \bibinfo{journal}{Phys. Rev. Lett.} \textbf{\bibinfo{volume}{109}},
  \bibinfo{pages}{096803} (\bibinfo{year}{2012}).

\bibitem[{\citenamefont{Pfeffer and Zawadzki}(2005)}]{q34}
\bibinfo{author}{\bibfnamefont{P.}~\bibnamefont{Pfeffer}} \bibnamefont{and}
  \bibinfo{author}{\bibfnamefont{W.}~\bibnamefont{Zawadzki}},
  \bibinfo{journal}{Phys. Rev. B} \textbf{\bibinfo{volume}{72}},
  \bibinfo{pages}{035325} (\bibinfo{year}{2005}).

\bibitem[{\citenamefont{Pfeffer and Zawadzki}(2006)}]{q36}
\bibinfo{author}{\bibfnamefont{P.}~\bibnamefont{Pfeffer}} \bibnamefont{and}
  \bibinfo{author}{\bibfnamefont{W.}~\bibnamefont{Zawadzki}},
  \bibinfo{journal}{Phys. Rev. B} \textbf{\bibinfo{volume}{74}},
  \bibinfo{pages}{233303} (\bibinfo{year}{2006}).

\bibitem[{\citenamefont{Dresselhaus}(1955)}]{q37}
\bibinfo{author}{\bibfnamefont{G.}~\bibnamefont{Dresselhaus}},
  \bibinfo{journal}{Phys. Rev.} \textbf{\bibinfo{volume}{100}},
  \bibinfo{pages}{580} (\bibinfo{year}{1955}).

\bibitem[{\citenamefont{Miller et~al.}(2003)\citenamefont{Miller, Zumb\"uhl,
  Marcus, Lyanda-Geller, Goldhaber-Gordon, Campman, and Gossard}}]{q38}
\bibinfo{author}{\bibfnamefont{J.~B.} \bibnamefont{Miller}},
  \bibinfo{author}{\bibfnamefont{D.~M.} \bibnamefont{Zumb\"uhl}},
  \bibinfo{author}{\bibfnamefont{C.~M.} \bibnamefont{Marcus}},
  \bibinfo{author}{\bibfnamefont{Y.~B.} \bibnamefont{Lyanda-Geller}},
  \bibinfo{author}{\bibfnamefont{D.}~\bibnamefont{Goldhaber-Gordon}},
  \bibinfo{author}{\bibfnamefont{K.}~\bibnamefont{Campman}}, \bibnamefont{and}
  \bibinfo{author}{\bibfnamefont{A.~C.} \bibnamefont{Gossard}},
  \bibinfo{journal}{Phys. Rev. Lett.} \textbf{\bibinfo{volume}{90}},
  \bibinfo{pages}{076807} (\bibinfo{year}{2003}).

\bibitem[{\citenamefont{Giglberger et~al.}(2007)\citenamefont{Giglberger,
  Golub, Bel'kov, Danilov, Schuh, Gerl, Rohlfing, Stahl, Wegscheider, Weiss
  et~al.}}]{q39}
\bibinfo{author}{\bibfnamefont{S.}~\bibnamefont{Giglberger}},
  \bibinfo{author}{\bibfnamefont{L.~E.} \bibnamefont{Golub}},
  \bibinfo{author}{\bibfnamefont{V.~V.} \bibnamefont{Bel'kov}},
  \bibinfo{author}{\bibfnamefont{S.~N.} \bibnamefont{Danilov}},
  \bibinfo{author}{\bibfnamefont{D.}~\bibnamefont{Schuh}},
  \bibinfo{author}{\bibfnamefont{C.}~\bibnamefont{Gerl}},
  \bibinfo{author}{\bibfnamefont{F.}~\bibnamefont{Rohlfing}},
  \bibinfo{author}{\bibfnamefont{J.}~\bibnamefont{Stahl}},
  \bibinfo{author}{\bibfnamefont{W.}~\bibnamefont{Wegscheider}},
  \bibinfo{author}{\bibfnamefont{D.}~\bibnamefont{Weiss}},
  \bibnamefont{et~al.}, \bibinfo{journal}{Phys. Rev. B}
  \textbf{\bibinfo{volume}{75}}, \bibinfo{pages}{035327}
  (\bibinfo{year}{2007}).

\bibitem[{\citenamefont{Studer et~al.}(2009{\natexlab{b}})\citenamefont{Studer,
  Salis, Ensslin, Driscoll, and Gossard}}]{q40}
\bibinfo{author}{\bibfnamefont{M.}~\bibnamefont{Studer}},
  \bibinfo{author}{\bibfnamefont{G.}~\bibnamefont{Salis}},
  \bibinfo{author}{\bibfnamefont{K.}~\bibnamefont{Ensslin}},
  \bibinfo{author}{\bibfnamefont{D.~C.} \bibnamefont{Driscoll}},
  \bibnamefont{and} \bibinfo{author}{\bibfnamefont{A.~C.}
  \bibnamefont{Gossard}}, \bibinfo{journal}{Phys. Rev. Lett.}
  \textbf{\bibinfo{volume}{103}}, \bibinfo{pages}{027201}
  (\bibinfo{year}{2009}{\natexlab{b}}).

\bibitem[{\citenamefont{Pfeffer and Zawadzki}(2012)}]{q41}
\bibinfo{author}{\bibfnamefont{P.}~\bibnamefont{Pfeffer}} \bibnamefont{and}
  \bibinfo{author}{\bibfnamefont{W.}~\bibnamefont{Zawadzki}},
  \bibinfo{journal}{J. Appl. Phys.} \textbf{\bibinfo{volume}{111}},
  \bibinfo{eid}{083705} (\bibinfo{year}{2012}).

\bibitem[{\citenamefont{Bychkov and Sheka}(1961)}]{q61}
\bibinfo{author}{\bibfnamefont{E.~I.} \bibnamefont{Bychkov}} \bibnamefont{and}
  \bibinfo{author}{\bibfnamefont{V.~I.} \bibnamefont{Sheka}},
  \bibinfo{journal}{Sov. Phys. Solid State} \textbf{\bibinfo{volume}{3}},
  \bibinfo{pages}{1257} (\bibinfo{year}{1961}).

\bibitem[{\citenamefont{Rashba and Sheka}(North-Holland, Amsterdam,
  1991)}]{q62}
\bibinfo{author}{\bibfnamefont{E.~I.} \bibnamefont{Rashba}} \bibnamefont{and}
  \bibinfo{author}{\bibfnamefont{V.~I.} \bibnamefont{Sheka}},
  \bibinfo{journal}{in \emph{Landau Level Spectroscopy}, p. 131, edited by G.
  Landwehr and E. I. Rashba}  (\bibinfo{year}{North-Holland, Amsterdam, 1991}).

\bibitem[{\citenamefont{Efros and Rashba}(2006)}]{q63}
\bibinfo{author}{\bibfnamefont{A.~L.} \bibnamefont{Efros}} \bibnamefont{and}
  \bibinfo{author}{\bibfnamefont{E.~I.} \bibnamefont{Rashba}},
  \bibinfo{journal}{Phys. Rev. B} \textbf{\bibinfo{volume}{73}},
  \bibinfo{pages}{165325} (\bibinfo{year}{2006}).

\bibitem[{\citenamefont{Shen et~al.}(2004)\citenamefont{Shen, Ma, Xie, and
  Zhang}}]{q42}
\bibinfo{author}{\bibfnamefont{S.-Q.} \bibnamefont{Shen}},
  \bibinfo{author}{\bibfnamefont{M.}~\bibnamefont{Ma}},
  \bibinfo{author}{\bibfnamefont{X.~C.} \bibnamefont{Xie}}, \bibnamefont{and}
  \bibinfo{author}{\bibfnamefont{F.~C.} \bibnamefont{Zhang}},
  \bibinfo{journal}{Phys. Rev. Lett.} \textbf{\bibinfo{volume}{92}},
  \bibinfo{pages}{256603} (\bibinfo{year}{2004}).

\bibitem[{\citenamefont{Das et~al.}(1990)\citenamefont{Das, Datta, and
  Reifenberger}}]{q43}
\bibinfo{author}{\bibfnamefont{B.}~\bibnamefont{Das}},
  \bibinfo{author}{\bibfnamefont{S.}~\bibnamefont{Datta}}, \bibnamefont{and}
  \bibinfo{author}{\bibfnamefont{R.}~\bibnamefont{Reifenberger}},
  \bibinfo{journal}{Phys. Rev. B} \textbf{\bibinfo{volume}{41}},
  \bibinfo{pages}{8278} (\bibinfo{year}{1990}).

\bibitem[{\citenamefont{Van'kov et~al.}(2006)\citenamefont{Van'kov, Kulik,
  Kukushkin, Kirpichev, Dickmann, Zhilin, Smet, von Klitzing, and
  Wegscheider}}]{q44}
\bibinfo{author}{\bibfnamefont{A.~B.} \bibnamefont{Van'kov}},
  \bibinfo{author}{\bibfnamefont{L.~V.} \bibnamefont{Kulik}},
  \bibinfo{author}{\bibfnamefont{I.~V.} \bibnamefont{Kukushkin}},
  \bibinfo{author}{\bibfnamefont{V.~E.} \bibnamefont{Kirpichev}},
  \bibinfo{author}{\bibfnamefont{S.}~\bibnamefont{Dickmann}},
  \bibinfo{author}{\bibfnamefont{V.~M.} \bibnamefont{Zhilin}},
  \bibinfo{author}{\bibfnamefont{J.~H.} \bibnamefont{Smet}},
  \bibinfo{author}{\bibfnamefont{K.}~\bibnamefont{von Klitzing}},
  \bibnamefont{and}
  \bibinfo{author}{\bibfnamefont{W.}~\bibnamefont{Wegscheider}},
  \bibinfo{journal}{Phys. Rev. Lett.} \textbf{\bibinfo{volume}{97}},
  \bibinfo{pages}{246801} (\bibinfo{year}{2006}).

\bibitem[{\citenamefont{Ando and Uemura}(1974)}]{q45}
\bibinfo{author}{\bibfnamefont{T.}~\bibnamefont{Ando}} \bibnamefont{and}
  \bibinfo{author}{\bibfnamefont{Y.}~\bibnamefont{Uemura}},
  \bibinfo{journal}{J. Phys. Soc. Jpn.} \textbf{\bibinfo{volume}{37}},
  \bibinfo{pages}{1044} (\bibinfo{year}{1974}).

\bibitem[{\citenamefont{Kallin and Halperin}(1984)}]{q46}
\bibinfo{author}{\bibfnamefont{C.}~\bibnamefont{Kallin}} \bibnamefont{and}
  \bibinfo{author}{\bibfnamefont{B.~I.} \bibnamefont{Halperin}},
  \bibinfo{journal}{Phys. Rev. B} \textbf{\bibinfo{volume}{30}},
  \bibinfo{pages}{5655} (\bibinfo{year}{1984}).

\bibitem[{\citenamefont{Dickmann and Kukushkin}(2005)}]{q47}
\bibinfo{author}{\bibfnamefont{S.}~\bibnamefont{Dickmann}} \bibnamefont{and}
  \bibinfo{author}{\bibfnamefont{I.~V.} \bibnamefont{Kukushkin}},
  \bibinfo{journal}{Phys. Rev. B} \textbf{\bibinfo{volume}{71}},
  \bibinfo{pages}{241310} (\bibinfo{year}{2005}).

\bibitem[{\citenamefont{Dickmann et~al.}(2005)\citenamefont{Dickmann, Zhilin,
  and Kulakovskii}}]{q48}
\bibinfo{author}{\bibfnamefont{S.}~\bibnamefont{Dickmann}},
  \bibinfo{author}{\bibfnamefont{V.}~\bibnamefont{Zhilin}}, \bibnamefont{and}
  \bibinfo{author}{\bibfnamefont{D.}~\bibnamefont{Kulakovskii}},
  \bibinfo{journal}{JETP} \textbf{\bibinfo{volume}{101}}, \bibinfo{pages}{892}
  (\bibinfo{year}{2005}).

\bibitem[{\citenamefont{Dickmann and Ziman}(2012)}]{q49}
\bibinfo{author}{\bibfnamefont{S.}~\bibnamefont{Dickmann}} \bibnamefont{and}
  \bibinfo{author}{\bibfnamefont{T.}~\bibnamefont{Ziman}},
  \bibinfo{journal}{Phys. Rev. B} \textbf{\bibinfo{volume}{85}},
  \bibinfo{pages}{045318} (\bibinfo{year}{2012}).

\bibitem[{\citenamefont{Bychkov and Martinez}(2002)}]{q50}
\bibinfo{author}{\bibfnamefont{Y.~A.} \bibnamefont{Bychkov}} \bibnamefont{and}
  \bibinfo{author}{\bibfnamefont{G.}~\bibnamefont{Martinez}},
  \bibinfo{journal}{Phys. Rev. B} \textbf{\bibinfo{volume}{66}},
  \bibinfo{pages}{193312} (\bibinfo{year}{2002}).

\bibitem[{\citenamefont{Bychkov et~al.}(1981)\citenamefont{Bychkov, Iordanskii,
  and M.}}]{q51}
\bibinfo{author}{\bibfnamefont{Y.~A.} \bibnamefont{Bychkov}},
  \bibinfo{author}{\bibfnamefont{S.~V.} \bibnamefont{Iordanskii}},
  \bibnamefont{and} \bibinfo{author}{\bibfnamefont{E.~G.} \bibnamefont{M.}},
  \bibinfo{journal}{JETP Letters} \textbf{\bibinfo{volume}{33}},
  \bibinfo{pages}{143} (\bibinfo{year}{1981}).

\bibitem[{\citenamefont{Antoniou and MacDonald}(1991)}]{q52}
\bibinfo{author}{\bibfnamefont{D.}~\bibnamefont{Antoniou}} \bibnamefont{and}
  \bibinfo{author}{\bibfnamefont{A.~H.} \bibnamefont{MacDonald}},
  \bibinfo{journal}{Phys. Rev. B} \textbf{\bibinfo{volume}{43}},
  \bibinfo{pages}{11686} (\bibinfo{year}{1991}).

\bibitem[{\citenamefont{Laughlin}(1983)}]{q64}
\bibinfo{author}{\bibfnamefont{R.~B.} \bibnamefont{Laughlin}},
  \bibinfo{journal}{Phys. Rev. Lett.} \textbf{\bibinfo{volume}{50}},
  \bibinfo{pages}{1395} (\bibinfo{year}{1983}).

\bibitem[{\citenamefont{Girvin et~al.}(1985)\citenamefont{Girvin, MacDonald,
  and Platzman}}]{q65}
\bibinfo{author}{\bibfnamefont{S.~M.} \bibnamefont{Girvin}},
  \bibinfo{author}{\bibfnamefont{A.~H.} \bibnamefont{MacDonald}},
  \bibnamefont{and} \bibinfo{author}{\bibfnamefont{P.~M.}
  \bibnamefont{Platzman}}, \bibinfo{journal}{Phys. Rev. Lett.}
  \textbf{\bibinfo{volume}{54}}, \bibinfo{pages}{581} (\bibinfo{year}{1985}).

\bibitem[{\citenamefont{Girvin et~al.}(1986)\citenamefont{Girvin, MacDonald,
  and Platzman}}]{q66}
\bibinfo{author}{\bibfnamefont{S.~M.} \bibnamefont{Girvin}},
  \bibinfo{author}{\bibfnamefont{A.~H.} \bibnamefont{MacDonald}},
  \bibnamefont{and} \bibinfo{author}{\bibfnamefont{P.~M.}
  \bibnamefont{Platzman}}, \bibinfo{journal}{Phys. Rev. B}
  \textbf{\bibinfo{volume}{33}}, \bibinfo{pages}{2481} (\bibinfo{year}{1986}).

\bibitem[{\citenamefont{Nefyodov et~al.}(2010)\citenamefont{Nefyodov,
  Fortunatov, Shchepetilnikov, and Kukushkin}}]{q53}
\bibinfo{author}{\bibfnamefont{Y.~A.} \bibnamefont{Nefyodov}},
  \bibinfo{author}{\bibfnamefont{A.~A.} \bibnamefont{Fortunatov}},
  \bibinfo{author}{\bibfnamefont{A.~V.} \bibnamefont{Shchepetilnikov}},
  \bibnamefont{and} \bibinfo{author}{\bibfnamefont{I.~V.}
  \bibnamefont{Kukushkin}}, \bibinfo{journal}{JETP Letters}
  \textbf{\bibinfo{volume}{91}}, \bibinfo{pages}{357–360}
  (\bibinfo{year}{2010}).

\bibitem[{\citenamefont{Nefyodov
  et~al.}(2011{\natexlab{a}})\citenamefont{Nefyodov, Shchepetilnikov,
  Kukushkin, Dietsche, and Schmult}}]{q54}
\bibinfo{author}{\bibfnamefont{Y.~A.} \bibnamefont{Nefyodov}},
  \bibinfo{author}{\bibfnamefont{A.~V.} \bibnamefont{Shchepetilnikov}},
  \bibinfo{author}{\bibfnamefont{I.~V.} \bibnamefont{Kukushkin}},
  \bibinfo{author}{\bibfnamefont{W.}~\bibnamefont{Dietsche}}, \bibnamefont{and}
  \bibinfo{author}{\bibfnamefont{S.}~\bibnamefont{Schmult}},
  \bibinfo{journal}{Phys. Rev. B} \textbf{\bibinfo{volume}{83}},
  \bibinfo{pages}{041307} (\bibinfo{year}{2011}{\natexlab{a}}).

\bibitem[{\citenamefont{Nefyodov
  et~al.}(2011{\natexlab{b}})\citenamefont{Nefyodov, Shchepetilnikov,
  Kukushkin, Dietsche, and Schmult}}]{q55}
\bibinfo{author}{\bibfnamefont{Y.~A.} \bibnamefont{Nefyodov}},
  \bibinfo{author}{\bibfnamefont{A.~V.} \bibnamefont{Shchepetilnikov}},
  \bibinfo{author}{\bibfnamefont{I.~V.} \bibnamefont{Kukushkin}},
  \bibinfo{author}{\bibfnamefont{W.}~\bibnamefont{Dietsche}}, \bibnamefont{and}
  \bibinfo{author}{\bibfnamefont{S.}~\bibnamefont{Schmult}},
  \bibinfo{journal}{Phys. Rev. B} \textbf{\bibinfo{volume}{84}},
  \bibinfo{pages}{233302} (\bibinfo{year}{2011}{\natexlab{b}}).

\bibitem[{\citenamefont{H\"ubner et~al.}(2011)\citenamefont{H\"ubner, Kunz,
  Oertel, Schuh, Pochwa\l{}a, Duc, F\"orstner, Meier, and Oestreich}}]{q56}
\bibinfo{author}{\bibfnamefont{J.}~\bibnamefont{H\"ubner}},
  \bibinfo{author}{\bibfnamefont{S.}~\bibnamefont{Kunz}},
  \bibinfo{author}{\bibfnamefont{S.}~\bibnamefont{Oertel}},
  \bibinfo{author}{\bibfnamefont{D.}~\bibnamefont{Schuh}},
  \bibinfo{author}{\bibfnamefont{M.}~\bibnamefont{Pochwa\l{}a}},
  \bibinfo{author}{\bibfnamefont{H.~T.} \bibnamefont{Duc}},
  \bibinfo{author}{\bibfnamefont{J.}~\bibnamefont{F\"orstner}},
  \bibinfo{author}{\bibfnamefont{T.}~\bibnamefont{Meier}}, \bibnamefont{and}
  \bibinfo{author}{\bibfnamefont{M.}~\bibnamefont{Oestreich}},
  \bibinfo{journal}{Phys. Rev. B} \textbf{\bibinfo{volume}{84}},
  \bibinfo{pages}{041301} (\bibinfo{year}{2011}).

\bibitem[{\citenamefont{Duckheim and Loss}(2006)}]{q58}
\bibinfo{author}{\bibfnamefont{M.}~\bibnamefont{Duckheim}} \bibnamefont{and}
  \bibinfo{author}{\bibfnamefont{D.}~\bibnamefont{Loss}},
  \bibinfo{journal}{Nature Physics} \textbf{\bibinfo{volume}{2}},
  \bibinfo{pages}{195} (\bibinfo{year}{2006}).

\bibitem[{\citenamefont{Kozuka et~al.}(2013)\citenamefont{Kozuka, Teraoka,
  Falson, Oiwa, Tsukazaki, Tarucha, and Kawasaki}}]{q59}
\bibinfo{author}{\bibfnamefont{Y.}~\bibnamefont{Kozuka}},
  \bibinfo{author}{\bibfnamefont{S.}~\bibnamefont{Teraoka}},
  \bibinfo{author}{\bibfnamefont{J.}~\bibnamefont{Falson}},
  \bibinfo{author}{\bibfnamefont{A.}~\bibnamefont{Oiwa}},
  \bibinfo{author}{\bibfnamefont{A.}~\bibnamefont{Tsukazaki}},
  \bibinfo{author}{\bibfnamefont{S.}~\bibnamefont{Tarucha}}, \bibnamefont{and}
  \bibinfo{author}{\bibfnamefont{M.}~\bibnamefont{Kawasaki}},
  \bibinfo{journal}{Phys. Rev. B} \textbf{\bibinfo{volume}{87}},
  \bibinfo{pages}{205411} (\bibinfo{year}{2013}).

\end{thebibliography}
\end{document}